\documentclass[aps,pra,preprint,showpacs, groupedaddress, doublespace]{revtex4}

\usepackage{braket}

\usepackage{graphics}

\usepackage{graphicx}

\usepackage{amssymb,amsmath}

\usepackage{setspace}

\DeclareMathAlphabet{\mathpzc}{OT1}{pzc}{m}{it}

\begin{document}






\preprint{}

%\setstretch{2}


\title{Preparation and detection of states with simultaneous spin alignment and molecular orientation in PbO}

\author{S. Bickman}
%\email{sarah.bickman@nist.gov}
%\affiliation{NIST, 325 Broadway, Boulder, CO 80305}
\altaffiliation{present address: NIST, 325 Broadway, Boulder, CO 80305}

\author{P. Hamilton}

\author{Y. Jiang}

\author{D. DeMille}
%\email{david.demille@yale.edu}

\affiliation{Yale University, PO Box 208120, New Haven, CT 06511, USA}

\date{\today}

\begin{abstract}

We are pursuing an experiment to measure the electric dipole moment of the electron using the molecule PbO.  This measurement requires the ability to prepare quantum states with orientation of the molecular axis and, simultaneously, alignment of the electron spin perpendicular to this axis.  It also requires efficient detection of the evolution of the spin alignment direction within such a state.  We describe a series of experiments that have achieved these goals, and the features and limitations of the techniques.  We also discuss possible new approaches for improved efficiency in this and similar systems.

\end{abstract}




\maketitle



\section{Introduction}

%\subsection{why edm is important}

A permanent electric dipole moment (EDM) of the electron, $\mathbf{d}_e$, would violate both parity and time-reversal invariance~\cite{CPstrange}, since $\mathbf{d}_e = 2d_e\mathbf{S}$ (where $\mathbf{S}$ is the electron spin).  There is substantial interest in measurements of $d_e$ with sensitivity beyond the current limit $d_e < 1.6 \times 10^{-27}$ $e \cdot$cm~\cite{thallium}.  A non-zero value of $d_e$ within the next few orders of magnitude would be a clear indication of physics beyond the Standard Model (SM).  Moreover, a non-zero EDM within this range is predicted to occur in a wide range of theories that extend the SM ~\cite{ComminsLeptons}.

%\subsection{why polar molecules}
Heavy polar molecules have a significant advantage over atoms in the search for $d_e$~\cite{sandars_measurability_1967}.  As a result of the strong hybridization of the atomic orbitals in such molecules, there is a large internal electric field $\hbox{\boldmath{$\mathcal{E}$}}_{int} = \mathcal{E}_{int} \hat{n}$, where $\hat{n}$ is the direction of the molecule's internuclear axis.  If $d_e\neq 0$, unpaired electrons interact with this internal field giving rise to a linear Stark shift described by the Hamiltonian $H_{EDM}=-\mathbf{d}_e\cdot\hbox{\boldmath{$\mathcal{E}$}}_{int}$.  This leads to an observable energy shift in molecules that can be 2-3 orders of magnitude larger than what is achievable in atoms under typical laboratory conditions~\cite{Sushkov,kozlov_parity_1995}.

Despite this advantage, molecules pose significant experimental challenges.  For example, the thermal Boltzmann distribution limits the population in any particular rovibrational state.  This reduces the counting rate in the experiment, and hence the overall sensitivity to $\mathbf{d}_e$.  In addition, molecules suitable for measuring $\mathbf{d}_e$ must have unpaired electron spins; such species are chemical free radicals and generally are not thermodynamically stable.  This leads to substantial experimental difficulties and typically reduces counting rates even further.

%\subsection{why PbO}

Measurements using the molecule PbO hold considerable promise for improved sensitivity to $\mathbf{d}_e$, as described in Refs. \cite{firstPbO, ComminsFetsch,2004} and summarized here.  PbO is not a free radical; rather, its ground state X(0)$^1\Sigma^+$  has closed shells ~\cite{Herzbergconstants}. The EDM measurement is conducted in the metastable a(1) $[^3\Sigma^+]$ state, which has two unpaired electron spins and hence is sensitive to $\mathbf{d}_e$.  The a(1) state can be populated by laser excitation and has a lifetime of $\tau_a \cong 82~\mu$s.  The $\Omega$-doublet substructure of this $| \Omega | = 1$ state allows it to be easily polarized with a modest external electric field $\mathcal{E} \gtrsim 15$ V/cm.
%The electronic angular momentum $\mathbf{J}_e$ can have a projection $\Omega$ %along or against the internuclear axis $n_z$, and this degeneracy is broken by the %Coriolis coupling between the electronic and rotational angular momentum.
Together these properties allow an EDM measurement using PbO contained in a closed cell, operating at substantial vapor density. These properties can in principle lead to an increase in the counting rate, over that available in molecular beam experiments, by as much as 7 orders of magnitude.  The primary topic of this paper is the development of methods for achieving efficient state preparation and detection, which are needed in order to take full advantage of these unique properties of PbO.

%\subsection{structure of a(1) state}

\begin{figure}
\includegraphics{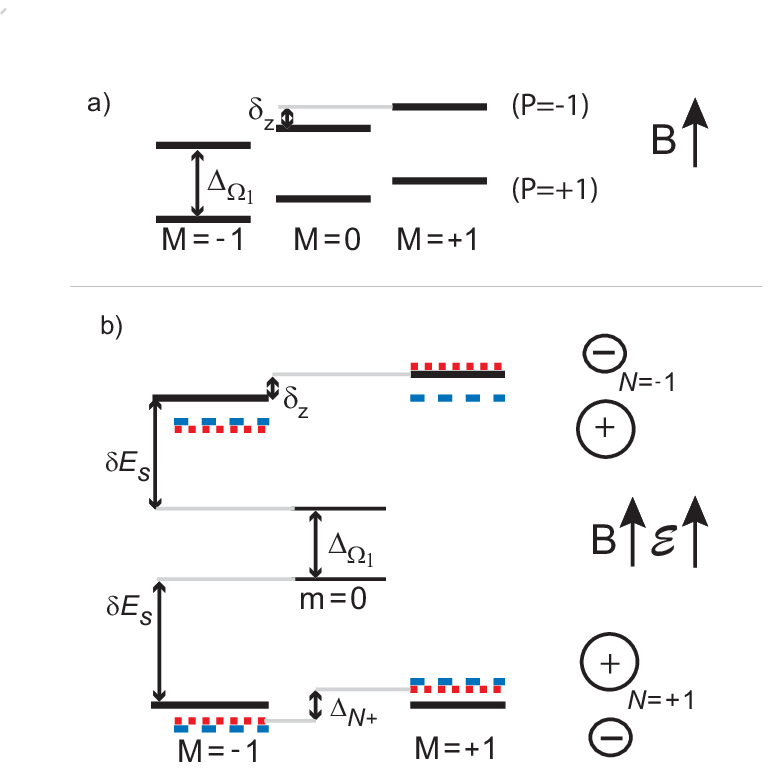}%
\caption{Energy level structure of the a(1) $J=1$ manifold. (a) Structure with no applied electric field ($\mathcal{E}=0$).  The eigenstates have well defined parity.  The $\Omega$-doublet splitting is $\Delta_{\Omega_1} \cong 2 \pi \times 11.214$ MHz~\cite{2004}. Zeeman shifts $\delta_Z$ due to the external magnetic field $B$ are shown.  (b) Structure with a fully polarizing electric field $\mathcal{E} \gg \Delta_{\Omega_1}/\mu_{\mathrm{a}}$.  Here sublevels with $|M| = 1$ are states of mixed parity.  Solid lines indicate the energy levels with only the Stark shift $\delta_{St}$ taken into account. (Typical values are $\delta_{St} \sim 2 \pi \times 40-100$ MHz in our experiments.) Dashed lines indicate levels with both Stark and Zeeman shifts $\delta_Z$ included.  (Typical values are $\delta_Z \sim 2 \pi \times 50-500$ kHz in our experiments.)  Dotted lines indicate the additional effects of a non-zero $\mathbf{d}_e$.
\label{fig:measscheme}}
\end{figure}

The $\Omega$-doublet structure of the a(1) state is central to the EDM measurement scheme using PbO, and introduces some new and unique aspects compared to previous EDM experiments.   Here we summarize the salient features of this structure.  Each a(1) vibrational state has a ladder of rotational levels with angular momentum $J=1, 2, ...$ and approximate energy $E_J = B_rJ(J+1)$, where the rotational constant $B_r \cong 2 \pi \times 7.054$ GHz \cite{Martin, Herzbergconstants} for the $v\!=\!5$ vibrational level used here.  (We take $\hbar\! =\! 1$ throughout.) Each rotational level contains a doublet of opposite parity states, separated by energy $\Delta_{\Omega_J} = qJ(J+1)$, where $q \ll B_r$.  The EDM measurement uses the $J\!=\!1$ manifold of states, whose energy structure is shown in Fig. \ref{fig:measscheme}a.  The $J\!=\!1$ doublet is separated by $\Delta_{\Omega_1} \cong 2 \pi \!\times\! 11.214$ MHz \cite{2004}.  These field-free states are designated by the ket $\ket{J,M,P,|\Omega |=1}$, where $M$ is the projection of $\mathbf{J}$ along the laboratory $z$-axis, $P = \pm 1$ is the parity, and $\Omega$ is the projection of the electronic angular momentum $\mathbf{J}_e$ along the molecular axis $\hat{n}$: $\Omega \equiv  \mathbf{J}_e \cdot \hat{n}$.  Alternatively, these states can be written in a basis with \textit{signed} values of $\Omega$, as $\ket{J,M,P,|\Omega |\! =\! 1} = \frac{1}{\sqrt{2}} \left\{ \ket{J,M,\Omega\! =\! +1} - (-1)^J P \ket{J,M,\Omega\! =\! -1} \right\}$ \cite{Brown2003}.  (Here and throughout the paper, kets without an electronic state label refer to the a(1) electronic state.)

For the EDM measurements, external fields must be applied. An external electric field $\hbox{\boldmath{$\mathcal{E}$}} = \mathcal{E}\hat{z}$ leads to an ordinary Stark interaction described by the Hamiltonian $H_{St} = -\hbox{\boldmath{$\mu$}}_a \cdot \hbox{\boldmath{$\mathcal{E}$}}$; here $\hbox{\boldmath{$\mu$}}_a = \mu_a \hat{n}$, where $\mu_a \cong 1.64$ MHz/(V/cm) is the molecule-fixed electric dipole moment in the a(1) state \cite{Hunter2002}.  Stark mixing of the $\Omega$-doublet states can be described by the Hamiltonian submatrix $H_{St}^{(J,M)}$ of the two-state subspace $\left( \ket{J,M,P=-1}, \ket{J,M,P=+1} \right)$:
\begin{equation}
H_{St}^{(J,M)} = \left(
\begin{array}{cc} %
\frac{\Delta_{\Omega_J}}{2} & ~~\frac{\mu_a \mathcal{E} M}{J(J+1)} \\
\frac{\mu_a \mathcal{E} M}{2}  & -\frac{\Delta_{\Omega_J}}{J(J+1)}
\end{array}
\right) \label{eq:2by2StarkHamiltonian}.
\end{equation}
These states have their energy shifted above (higher state $H$) or below (lower state $L$) the field-free values; the magnitude of the shift $\delta_{St}$ is given by 
\begin{equation}
\delta_{St} = \sqrt{\left(\frac{\Delta_{\Omega_J}}{2}\right)^2 + \frac{\mu_a^2 \mathcal{E}^2 M^2}{[J(J+1)]^2}} - \frac{\Delta_{\Omega_J}}{2}.
\end{equation}
The Stark mixing electrically polarizes the molecule, so that the resulting eigenstates have a nonzero expectation value of $\hbox{\boldmath{$\mu$}}_a$.  Defining the polarization direction as $N \equiv \mathrm{sign}(\langle\hbox{\boldmath{$\mu$}}_a\rangle \cdot \hbox{\boldmath{$\mathcal{E}$}})$, the resulting $L$ $(H)$ eigenstate has $N = +1$ ($N=-1$) for any finite value of $\mathcal{E}$.  
In the limit of large $\mathcal{E}$ (such that $\mu_a \mathcal{E} \gg \Delta_{\Omega_J}$), the $\Omega$-doublet states are \textit{fully} mixed; to describe these states it is convenient to define the new basis set $\ket{J,M,N} \equiv \ket{J,M,\Omega\! =\! N \cdot \mathrm{sign}(M)} = \frac{1}{\sqrt{2}}\left\{\ket{J,M,P\! =\! +1}-(-1)^{\Omega}\ket{J,M,P\! =\! -1}\right\}$.  These states have maximal polarization, $\langle\hbox{\boldmath{$\mu$}}_a\rangle = N \mu_a |M|/[J(J+1)]\hat{z}$.  Note that the $M\! =\! 0$ sublevels always remain unmixed, and that we have neglected mixing with much more distant states of different $J$.

The effect of a magnetic field $\mathbf{B} = B\hat{z}$ on the system can be described by the effective Hamiltonian $H_Z^{(\mathrm{eff})} = \mu_B g_{H,L}^{(J)}(\mathcal{E})\mathbf{J}\cdot\mathbf{B}$, where $\mu_B$ is the Bohr magneton.  This Hamiltonian leads to Zeeman shifts of sublevels with quantum number $M$ by $\delta_{Z_{H,L}} \equiv g_{H,L}^{(J)}(\mathcal{E}) \mu_B B M$ compared to the field-free energies.  The dependence of the $g$-factors on $\mathcal{E}$ is discussed in detail in section VI.  Regardless of the value of $\mathcal{E}$, roughly $g_{H}^{(J)} \cong g_{L}^{(J)} \cong \frac{G_\parallel}{J(J+1)}$ (here $G_\parallel \cong 1.86$ \cite{2004}), with differences at the level of a few parts per thousand.  The relevant $J\! =\! 1$ level structure in the presence of magnetic and electric fields is shown in Fig. \ref{fig:measscheme}.  We assign special symbols to replace the subscripts $H,L$ on $g$ and $\delta_Z$ in some limiting cases: $H (L) \rightarrow P_- (P_+)$ when $\mathcal{E} \! =\! 0$ (parity eigenstates), and $H (L) \rightarrow N_- (N_+)$ for large $\mathcal{E}$ (fully polarized states). Since the paper focuses mainly on $J\! =\! 1$ states, we omit the superscript $(J)$ for these states unless needed for clarity.  Also, we sometimes refer to a generic Zeeman splitting as $\delta_Z$, with no subscript.

The desired EDM measurement sequence starts by populating a coherent superposition of the $\ket{J\! =\! 1,M\! =\! \pm 1,N\! =\! +1}$ states and measuring the energy difference $\Delta_{N_+}$ between these two levels; or by populating the $\ket{J\! =\! 1,M\! =\! \pm 1,N\! =\! -1}$ states and measuring their difference $\Delta_{N_-}$. Because of the different signs of $\left\langle \hbox{\boldmath{$\mathcal{E}$}}_{int} \right\rangle = \mathcal{E}_{int} N \hat{z}$ in these two states, the effect of $\mathbf{d}_e$ is of opposite sign in these two measurements, while the Zeeman shifts are nearly identical.  Explicitly, we write $\Delta_{N\pm} = \pm 2 d_e \mathcal{E}^{(eff)}_{int} + 2 g_{N\pm} \mu_B B$. Here $\mathcal{E}^{(eff)}_{int} \approx 26$ GV/cm \cite{Titov,Meyer2006} is the effective internal field acting on the EDMs of the pair of valence electrons in the a(1) state.  If $g_{N_+} = g_{N_-}$ exactly, the quantity $\Delta_{EDM} = \Delta_{N_+} - \Delta_{N_-} = 4 d_e \mathcal{E}^{(eff)}_{int}$ is entirely independent of $B$.  Hence, comparison of the energy shifts in the oppositely-polarized levels (with $N \! =\!  \pm 1$) allows for the cancellation both of magnetic noise sources and of systematic effects due to magnetic fields; i.e., the $\Omega$-doublet structure acts as an internal comagnetometer~\cite{ComminsFetsch}. %Finally, consider the effect of reversing the sign of $\hbox{\boldmath{$\mathcal{E}$}}$ with respect to %$\mathbf{B}$.  In this case, the sign of $N$ is switched between the upper and lower energy levels.  has Note %that here the effect is $\Delta_{N\pm}$ and in $\delta_{EDM}$.

Suppose we initially prepare a particular initial state such as $\ket{\psi(T\! =\! 0)} = \frac{1}{\sqrt{2}}(\ket{J\! =\! 1,M\! =\! +1,N\! =\! +1}+\ket{J\! =\! 1,M\! =\! -1,N\! =\! +1})$.  After free evolution for time $T$, the state will evolve to $\ket{\psi(T)} = \frac{1}{\sqrt{2}}(e^{-i \theta/2}\ket{J\! =\! 1,M\! =\! +1,N\! =\! +1}+e^{i \theta/2}\ket{J\! =\! 1,M\! =\! -1,N\! =\! +1})$, where $\theta = \Delta_{N_+} T$.  Measurement of $\Delta_{N_+}$ is thus equivalent to measuring the time-dependent phase $\theta$.  Note that the state $\ket{\psi(T)}$ has, simultaneously, a well-defined molecular orientation (corresponding to the sign of $N$) and a well-defined spin alignment perpendicular to the molecular axis.  For the latter, we mean specifically that the expectation values $\left\langle J^2_x \right\rangle \propto \cos^2{\theta/2}$ and $\left\langle J^2_y \right\rangle \propto \sin^2{\theta/2}$, while $\left\langle J_x \right\rangle = \left\langle J_y \right\rangle = \left\langle J_z \right\rangle = 0$.  (For precise definitions of orientation and alignment, see e.g. Ref. \cite{Budker2004}.)  In this sense, the evolution of $\ket{\psi(T)}$ can be described as a precession of the spin alignment about the $z$-axis, and measurement of $\Delta_{N_+}$ can be achieved by monitoring the alignment direction.

In order to implement the desired measurement, it has been necessary to devise new techniques to prepare this unusual type of state.  This requires methods beyond simple laser excitation, since the molecule-oriented states (with definite value of $N$) are not resolved in the presence of usual Doppler broadening.  Moreover, these states are of mixed parity so that standard selection rules are modified.  Similarly, the detection of $\Delta_{N\pm}$ also requires sub-Doppler methods, since under typical conditions in our work $\Delta_{N\pm} \sim 50-1000$ kHz (determined by the applied field $B \sim 0.025-0.5$ G). Our detection methods explicitly take advantage of the spin alignment precession, to directly measure $\theta$.  Although our discussion centers on the specific case of the electron EDM measurement using PbO, we believe our techniques for preparing states with simultaneous molecular orientation and spin alignment may find application in other systems and for other purposes (e.g. for encoding of qubits into the spin degree of freedom of polar molecules) \cite{Andre2006}.

%\subsection{outline of paper}
The discussion will begin with an overview of the key features of the apparatus used for these measurements.  Next, simple schemes for state preparation and detection in the absence of an applied electric field, as used in our earlier work \cite{2004}, will be described.  In this context, the issues of population and detection efficiency relevant to the EDM measurement will be introduced.  Then, three different schemes suitable for preparation of molecule-oriented states will be discussed.  Finally, two additional detection schemes will be discussed along with their implications for improved sensitivity to $\mathbf{d}_e$.  On several occasions we use these techniques to obtain various types of spectroscopic data on PbO; these measurements are described in the text as is relevant.

\section{Apparatus}
\begin{figure}
\includegraphics[width=3.3in]{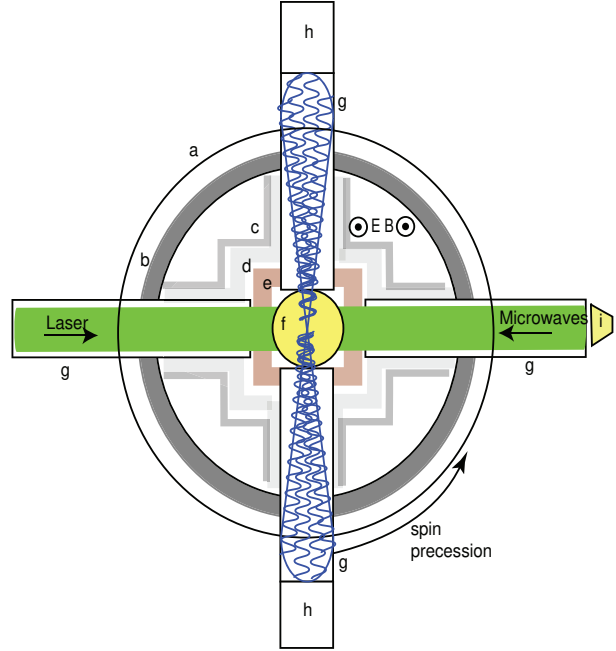}
\caption{(color online) Cross sectional top view of the apparatus.  The apparatus consists of: (a) up to four layers of nearly cylindrical magnetic shielding; (b) the aluminum vacuum chamber; (c) heat shields; (d) the quartz and tantalum oven; (e) the alumina and sapphire vapor cell; (f) reentrant gold electrodes; (g) quartz lightpipes; (h) fluorescence collection optics and detectors; and (i) microwave electronics and horn. 
\label{fig:apparatus}}
\end{figure}

%\subsection{ oven, cell}
The central apparatus for this experiment consists of a vapor cell consisting of an alumina rectangular frame, 2.5" tall with 3.5" wide sides (see Fig. \ref{fig:apparatus}).  The cell has 2" diam. holes covered by sapphire windows on four faces for optical access. The windows are held on the cell using gold foils as a sealant \cite{bailey}.  A stem tube, held at a temperature slightly below that of the main cell body, provides a reservoir of isotopically enriched $^{208}$PbO. A movable alumina plunger seals a hole in the top of the cell; this hole is used to evacuate the cell as needed. Electric fields are applied with gold circular electrodes (2" diam.) and annular guard rings (2.25" i.d., 2.5" o.d.), in a reentrant geometry to provide a suitable aspect ratio for achieving a uniform field (1.5" separation). Typically the guard ring voltage is $\approx 1.8$ times larger than the electrode voltage to achieve optimum field homogeneity. The electrodes and guard rings are separated from each other and from the main cell by high-purity sapphire and/or beryllia insulating spacers. The cell is radiatively heated to 700$^\circ$C with resistive tantalum heaters supported by an opaque quartz frame.  The cell temperature is measured with non-magnetic thermocouples, developed for this experiment, consisting of a Au/(50\%Au-50\%Pd) junction. The heaters, which are laid out to minimize their inductance, are driven with audio-frequency currents. The amplitude is modulated so that both drive currents and eddy currents are strongly suppressed during the measurement time after each laser pulse. The oven is surrounded by several layers of metal and quartz heat shields, and sits within an aluminum vacuum chamber.  2" diam. quartz rods are used as vacuum feedthroughs to convey light into and out of the oven to the cell.

%To collect laser-induced fluorescence from the PbO molecules, two of these quartz %rods are polished on their circumference so they act as light pipes; the fluorescent %light passes through various spectral filters and is detected using photomultiplier %tubes.

%\subsection{lasers}

A pulsed Nd:YAG-pumped dye laser (rep. rate 100 Hz, energy $\sim\! 15-25$ mJ/pulse) is used to populate the a(1) $\ket{J\! =\! 1}$ manifold of levels by driving the X$(v\! =\! 1, J\! =\! 0) \rightarrow$a(1)$(v\! =\! 5,J\! =\! 1)$ transition at wavelength $\lambda = 570$ nm.  The nominal laser propagation direction is defined as the $\hat{y}$ axis.  We observe that the time-averaged laser spectrum can be described as the sum of two peaks with the same center frequency: one $\sim 1.7$ GHz FWHM containing $\sim 1/3$ the power, and the other $\sim 12$ GHz FWHM containing the remainder of the power.
%The linewidth of this laser has two components: a 1.5 GHz FWHM containing most %of the power, and a 10 GHz FWHM containing the remaining power. [Can we put %any numbers on how much power is in what ranges of the laser linewidths? Or is %there a better way or expressing this?]
%\subsection{fluorescence detection}
The a(1) state molecules are detected by monitoring their decay fluorescence along the a(1)$(v=5)\rightarrow$X$(v=0)$ channel at $\lambda = 549$ nm. Fluorescence is collected along the $\pm\hat{x}$ direction by two of the quartz rods, which are polished on their circumference to act as light guides.  The fluorescent light passes through color glass and interference filters to block scattered laser light and blackbody radiation from the oven.  The transmitted light is detected using photomultiplier tubes (PMTs) which are gated to further reduce scattered light. The signals are digitized, stored, and analyzed using a PC-based data acquisition system.

\section{Preparation and detection of spin alignment \textit{without} molecular orientation}

\begin{figure}
\includegraphics[width=3.31 in]{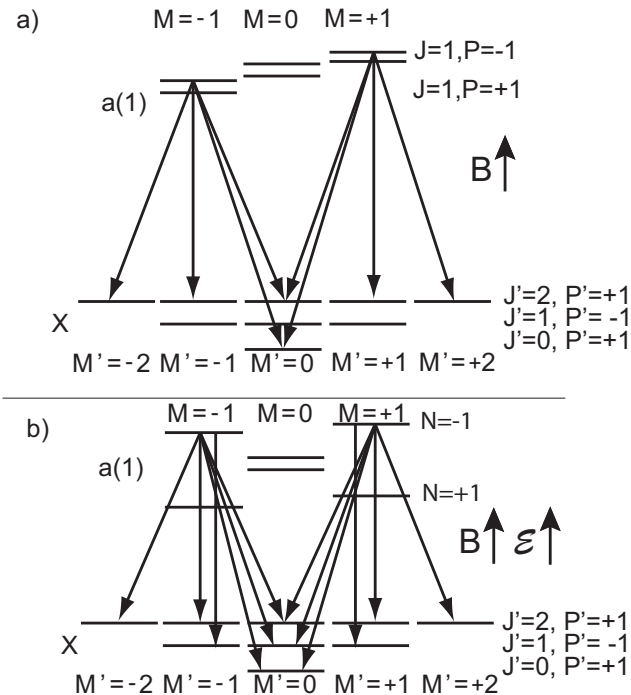}%
\caption{Allowed decay channels for the a(1)$J=1$ states.  a) Decays of the $\ket{J\! =\! 1,M\! =\! \pm 1,P\! =\! -1}$ levels.  Only final states with $P'\! =\! +1\! =\! (-1)^{J'}$ are allowed. b) Decays of oriented molecule states such as $\ket{J\! =\! 1,M\! =\! \pm 1, N=+1}$.  These mixed-parity states can decay to final states of both parities. In both cases, only decays to common ($M'=0$) sublevels give rise to quantum beats. \label{fig:decays}}
\end{figure}

We begin with a brief review of the technique used to prepare and detect spin alignment in the $\ket{\mathrm{a}, J\! =\! 1}$ levels, when no electric field is present \cite{2004}.  The laser pulse driving the $\ket{\mathrm{X},J\! =\! 1, P\! =\! +1} \rightarrow \ket{\mathrm{a},J\! =\! 1}$ transition is linearly polarized along $\hat{x}$.  Standard electric dipole selection rules then ensure population of the state
\begin{equation}
\ket{\psi(T=0)} = \ket{\psi_{P-}^{(0)}} \equiv \frac{1}{\sqrt{2}} (\ket{J\! =\! 1,M\! =\! +1,P\! =\! -1} - \ket{J\! =\! 1,M\! =\! -1,P\! =\! -1})
\end{equation}
if the laser pulse duration $\tau_L$ satisfies the condition $\tau_L \ll \delta_Z^{-1}$, so that the $M\! =\! \pm 1$ sublevels cannot be resolved.  This state evolves at time $T$ to $\ket{\psi(T)} = \ket{\psi_{P-}(T)} \equiv \frac{1}{\sqrt{2}} (e^{-i\delta_Z T}\ket{J\! =\! 1,M\! =\! +1,P\! =\! -1} - e^{i\delta_Z T}\ket{J\! =\! 1,M\! =\! -1,P\! =\! -1})$.  The measured fluorescence signal $S(T)$ arises from decay of this state to all possible sublevels $\ket{J',M',P'}$ of the X $^1\Sigma^+(v=0)$ state, where $P'\! = \! (-1)^{J'}$ for all levels.  Then
\begin{equation}
S(T)\propto \sum_f \left|\bra{\mathrm{X},J',M',P'} \hbox{\boldmath{$\epsilon$}}\cdot\mathbf{r}\ket{\psi (T)} \right|^2 ; \label{eq:quantumbeats1}
\end{equation}
here $f$ denotes all possible final states of both the molecule and the polarization of the outgoing photon $\hbox{\boldmath{$\epsilon$}}$.  For detection along the $x$-axis, $\hbox{\boldmath{$\epsilon$}} = \hat{y}$ or $\hat{z}$ spans the possibilities.  Moreover, electric dipole selection rules ensure that only $J'\! =\! 0$ and $J'\! =\! 2$ states contribute to the signal for $\ket{\psi(T)} = \ket{\psi_{P-}(T)}$ (see Fig. \ref{fig:decays}).  Hence
\begin{equation}
S(T)\propto \sum_{J'=0,2} \sum_{M'=-J'}^{J'} \sum_{j=y,z} \left| \bra{\mathrm{X},J',M',P'\! =\! +1}
r_j \ket{\psi_{P-}(T)} \right|^2. \label{eq:quantumbeats}
\end{equation}
Within this sum, decays to $M'=0$ states are of most interest since here interference gives rise to quantum beats;  hence we write $S(T) = S_{qb}(T) + S_{bg}(T)$. Here $S_{qb}~[S_{bg}]$ is the quantum beat [background] part of the signal, given by
\begin{equation}
S_{qb}(T) = \sum_{J'=0,2} \left| \bra{\mathrm{X},J',M'=0,P'\! =\! +1}
y \ket{\psi_{P-}(T)} \right|^2 = S_{qb}^{(0)}\left[1+\cos(2\delta_Z T)\right]
\end{equation}
and
\begin{equation}
S_{bg}(T) = \sum_{\stackrel{M'=-J'}{M' \neq 0}}^{J'} \sum_{j=y,z} \left| \bra{\mathrm{X},J'\! =\! 2,M',P'\! =\! +}
r_j \ket{\psi_{P-}(T)} \right|^2 = S_{bg}^{(0)}.
\end{equation}
Here we have used the fact that only $\hbox{\boldmath{$\epsilon$}} = \hat{y}$ polarization contributes to $\delta M\! =\! \pm 1$ transitions that are detected along $\hat{x}$.  To derive the quantum beat signal it is useful to express $y = -(r_{(+1)}\!  +\!  r_{(-1)})/\sqrt{2}$ (where $r_{(q)}$ are spherical tensor components) and use the Wigner-Eckhart theorem to relate the matrix elements $\bra{\mathrm{X},J',M',P'} r_j \ket{\mathrm{a},J,M,P}$ for different values of $M, M'$.  We emphasize that $2 \delta_Z$ is exactly the desired energy splitting between the $M\! =\! \pm 1$ components, and that the relative phase $\theta = 2 \delta_Z T$ corresponds to the precession angle of the spin alignment in the $x-y$ plane.

\begin{figure}
\includegraphics[width=3.31 in]{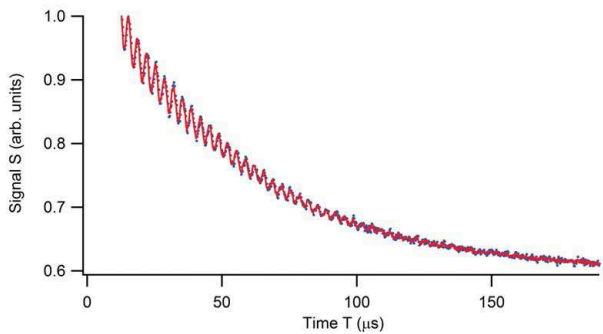}%
\caption{(color online) Typical Zeeman quantum beats, measured in the absence of an applied electric field ($\mathcal{E}=0)$.  Both data (blue points, average over 70 laser pulses) and the fit function described in the text (red line) are shown.  For this data, $c=9.2(1)\%$, $\omega_b = 2 \pi \times 299.8(1)$ kHz, and $\phi=5.02(1)$ radians. The background/signal ratio $d/\alpha \cong 1.10$ is evident from the vertical scale. \label{fig:quantumbeats}}
\end{figure}

We rewrite the signal as $S(T) \propto 1 + c\cos(2 \delta_Z T)$, where $c$ is the contrast of the quantum beats.  The intrinsic value $c = c^{(0)} = S_{qb}^{(0)}/(S_{qb}^{(0)}+S_{bg}^{(0)})$ is determined by the branching ratio for $\ket{\mathrm{a},J\! =\! 1, P\! =\! -1}$ decays to $\ket{\mathrm{X},J\! =\! 2, P\! =\! +1}$ versus $\ket{\mathrm{X},J\! =\! 0, P\! =\! +1}$ states.  For allowed transitions, this could be calculated from H$\ddot{\mathrm{o}}$nl-London factors, but for the forbidden X$[^1\Sigma^+] \rightarrow \textrm{a}~[^3\Sigma^+]$ transition no reliable calculation exists.  Based on our measurements of relative line strengths for different $J'\rightarrow J$ transitions, we estimate $c^{(0)} \approx 50\%$.  However, in practice there are other sources of background that reduce the observed contrast.  In our experiments the broad spectral width of the laser leads to excitation of higher-$J$ rotational levels that contribute to the fluorescence signal but not the quantum beats.  Under typical conditions, we find $c\approx$10\% as shown in Fig. \ref{fig:quantumbeats}.

%\subsection{fitting including scrambled data}
We fit our data to a function of the form
\begin{equation}
S(T) = \alpha I(T)\left[1 + c~e^{-T/T_b}\cos{(\omega_b T + \phi)}\right] + d.
\label{eq:fitequation}
\end{equation}
Here $I(T)$ gives the temporal evolution of the signal in the absence of quantum beats, normalized such that $I(0)=1$; typically $I(T) \approx e^{-T/T_1}$ (with $T_1 \approx 40~\mu$s determined by a combination of the radiative lifetime $\tau_a$ and collisional quenching of the state on cell walls).  The free parameters in the fit are the signal size $\alpha$; the quantum beat contrast $c$, frequency $\omega_b$, and phase $\phi$; the DC background $d$; and the beat shortening time $T_b$ (due to spin-decohering collisions).  At various points in the experiment, $I(T)$ was obtained by different methods such as a) deliberately applying an inhomogeneous magnetic field to rapidly dephase the beats; b) populating the $\ket{J\! =\! 1,M\! =\! 0,P\! = \! -1}$ state by setting the excitation laser polarization along $\hat{z}$; and c) filtering the quantum beat data using methods similar to those in Ref.~\cite{Duong,Paley}.  All methods give similar results and will not be distinguished from here on.  The background $d$ is due primarily to blackbody radiation from the oven.  Typically the first $\sim \! 10~\mu$s of data is not fit: this data is contaminated by transients associated with PMT gating, and by fluorescence from excitation of the short-lived A(0) $^3\Pi$ state by spectral wings of the laser.

\section{Statistical sensitivity to energy shifts}

%\subsecton{backgrounds and contrast}
Consider an EDM measurement using methods similar to those just described, but with an oriented molecular state prepared by some method.  Then the beat frequency is $\omega_b = \Delta_{N\pm}$, and the EDM uncertainty $\delta d_e = \delta \omega_b /\left(2 \mathcal{E}_{int}^{(eff)}\right)$.  Here $\delta \omega_b$, the uncertainty in the fitted value of $\omega_b$, determines the detectable size of $d_e$.  Hence, our goal is to minimize $\delta \omega_b$ for measurements on oriented molecular states.  A lower limit on $\delta \omega_b$ is set by the shot noise on the fluorescence signal.  The signal is $S(T) = G \dot{N}(T)$, where $\dot{N}(T)$ is the instantaneous photoelectron rate at the PMT cathode and $G$ is an electronic gain.  Straightforward analysis shows that the shot noise-limited uncertainty for a single laser pulse is
\begin{equation}
\delta \omega_b^{(s)} = \frac{\sqrt{2}}{c T_2 \sqrt{\dot{N}(0)T_2}}F(T_1/T_b, d/\alpha).
\label{eq:frequncertaintyformula}
\end{equation}
Here the beat coherence time $T_2$ is given by $T_2^{-1} = T_b^{-1} + T_1^{-1}$, and the function $F(T_1/T_b, d/\alpha) \cong \sqrt{(\frac{1+2T_1/T_b}{1+T_1/T_b})^3 + 8\frac{d}{\alpha}}$.  Under typical conditions $T_b \approx T_1$; thus for a fixed molecular signal $\propto \alpha$, $F \approx \sqrt{\frac{27}{8} + 8\frac{d}{\alpha}}$ and the frequency uncertainty degrades rapidly for background/signal ratio $d/\alpha \gtrsim 1/3$.  The factor of $\sqrt{2}$ in the numerator of Eq. \ref{eq:frequncertaintyformula} arises from treating the phase $\phi$ as a free parameter in the fit.  (This analysis neglects the effect of removing the first $\sim \! 10 \mu$s of data from the fit.)  Under typical conditions for the measurements discussed in sections III-VIII, $\dot{N} \approx 10^{10}$/s and $T_2 \approx 30~\mu$s.  Over the course of these experiments $d$ varied substantially, but typically $d/\alpha \approx 2-10$ and $F \approx 4-9$.  With $c \approx 10\%$, typically $\delta \omega_b \approx 2 \pi \times 250$ Hz under these conditions. In section X and Appendix A, the conditions determining $\dot{N}$ and $d$ are discussed in detail.  Our measured frequency uncertainty was typically within a factor of $1.0-1.3$ of the shot noise limit given by Eq. \ref{eq:frequncertaintyformula}.

\section{Quantum beats in fluorescence from oriented molecular states}

In our initial experiments with oriented molecular states, we again used quantum beats measured in fluorescence to determine $\Delta_{N\pm}$.  In these cases, the only parameter that could be optimized to increase sensitivity was the beat contrast $c$.  Our discussion in the next few sections thus concentrates on methods to improve $c$ when detecting oriented states.  In other experiments discussed later, we investigated alternate detection schemes in an attempt to further improve $c$ and, at the same time, to increase both the signal/background ratio and the absolute signal size.

Quantum beats in fluorescence from oriented molecular states such as
$\ket{\psi_{N_+}(T)}\equiv \left(e^{-i\Delta_{N_+}T/2}\ket{J\! =\! 1,M\! =\! +1,N\! =\! +1}\!-\!e^{+i\Delta_{N_+}T/2}\ket{J\! =\! 1,M\! =\! -1,N\! =\! +1}\right)/\sqrt{2}$ are similar to those discussed earlier.  However, the intrinsic contrast for these mixed-parity states is different than for the $P\! =\! -1$ state, since now decays to the states $\ket{\mathrm{X},J'\! =\! 1,M',P'\! =\! -1}$ are also allowed.  (This is the only molecular final state accessible from the $P\! =\! +1$ components of $\ket{\psi_{N_+}(T)}$.) A calculation similar to that above yields a slight reduction in expected intrinsic contrast: $c^{(0)} \approx 45\%$ for these polarized states.  As before the observed constrast is substantially reduced due to fluorescence from other rotational states, but taking this into account our observations are in rough qualitative agreement with this prediction.

\section{Laser Excitation With an Applied Electric Field}
%\subsection{horizontal laser}
The simplest method to prepare states with simultaneous molecular orientation and spin alignment is essentially identical to the technique described above, but with the external electric field $\mathcal{E}$ present throughout the excitation and detection.  A fully polarizing $\mathcal{E}$ causes all four $J\! =\! |M|\! =\! 1$ sublevels to be states of mixed parity, and the laser linewidth is significantly broader than the Stark shifted energy structure of the a(1) state; hence a $\hat{x}$-polarized laser will excite a coherent superposition of all four of these levels.  Initially, the excited state will be
\begin{eqnarray}
\ket{\psi_{\mathcal{E}}(T\! =\! 0)} &=& \ket{\psi_{P-}^{(0)}} \nonumber \\
& = & \frac{1}{2} ( \ket{J\! =\! 1,M\! =\! -1, N\! =\! +1}-\ket{J\! =\! 1,M\! =\! +1, N\! =\! +1}\nonumber\\
& ~ & ~ - \ket{J\! =\! 1,M\! =\! -1, N\! =\! -1}+\ket{J\! =\! 1,M\! =\! +1, N\! =\! -1} ).
\end{eqnarray}
In ideal homogeneous fields $\mathcal{E}$ and $B$, the combined Stark and Zeeman shifts will cause this state to evolve as
\begin{eqnarray}
&~& \ket{\psi_{\mathcal{E}}(T)} \nonumber\\
&=&  \frac{1}{2}(e^{-i(\omega_S - \delta_{Z_{N_+}})T}\ket{J\! =\! 1,M\! =\! -1,N\! =\! +1}
-e^{-i(\omega_S + \delta_{Z_{N_+}})T}\ket{J\! =\! 1,M\! =\! +1,N\! =\! +1} \nonumber \\
& -& e^{-i(-\omega_S - \delta_{Z_{N_-}})T}\ket{J\! =\! 1,M\! =\! -1,N\! =\! -1}
+e^{-i(-\omega_S + \delta_{Z_{N_-}})T}\ket{J\! =\! 1,M\! =\! +1,N\! =\! -1}).
\end{eqnarray}
Here $\omega_S \cong \delta_{St} + \Delta_{\Omega_1}/2$, and we ignore the small effect of the electron EDM.
Because $\omega_S \gg \delta_{Z_{N\pm}} \gg 1/T_2$, even the small electric field inhomogeneity in the system ($\delta \mathcal{E}/\mathcal{E} \lesssim 1\%$) leads to rapid decoherence between the $N=+1$ and $N=-1$ components of $\ket{\psi_{\mathcal{E}}(T)}$.  The resulting system can be described as an incoherent mixture of two states, each of which has simultaneous molecular orientation and spin alignment:
\begin{eqnarray}
\ket{\psi_{N_+}(T)}= \frac{e^{\!-i \delta_{Z_{N_+}}T}\ket{J\! =\! 1,M\! =\! -1,N\! =\! +1}\!-\!e^{+i\delta_{Z_{N_+}}T}\ket{J\! =\! 1,M\! =\! +1,N\! =\! +1}}{\sqrt{2}};\\
\ket{\psi_{N_-}(T)}=\frac{e^{\!-i \delta_{Z_{N_-}}T}\ket{J\! =\! 1,M\! =\! -1,N\! =\! -1}\!-\!e^{+i\delta_{Z_{N_-}}T}\ket{J\! =\! 1,M\! =\! +1,N\! =\! -1}}{\sqrt{2}}.
\end{eqnarray}

\begin{figure}
\includegraphics[width=3.3in]{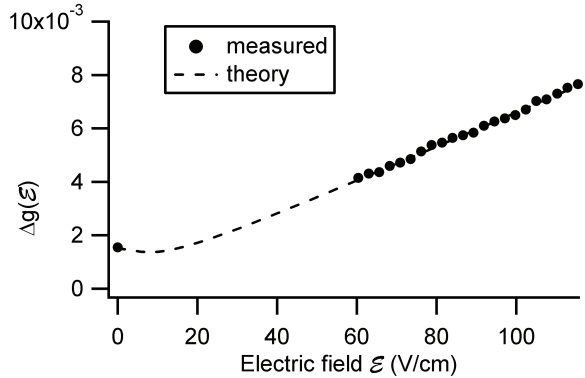}%
\caption{Plot of difference in g-factors $\Delta g$ between higher and lower energy $\Omega$-doublet states, as a function of applied electric field $\mathcal{E}$.  Points are data; the dashed line is the result of the analytic calculation discussed in the text.  The excellent agreement is evident. \label{fig:geff}}
\end{figure}

%\subsection{data with g(E)}
If $g_{N_+} = g_{N_-}$, quantum beats in fluorescence from these states are indistinguishable.  As described in our earlier work \cite{2004}, we initially expected this to be the case: the Stark-induced mixing of the $J\! =\! 1, P=\! \pm 1$ levels causes their $g$-factors (initially different by the small amount $g_{P_-}-g_{P_+} = 30(8) \times 10^{-4}$) to converge.  However, our data indicates that the $g$-factors of the polarized states actually \textit{diverge} as $\mathcal{E}$ increases (see Fig. \ref{fig:geff}).  This is explained below.

\begin{figure}
\includegraphics[width=3.3in]{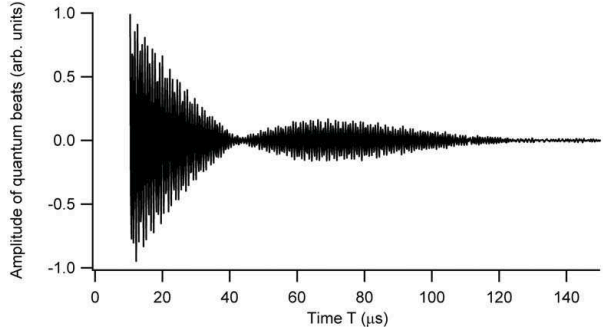}%
\caption{Quantum beats with an incoherent superposition of oriented molecular states.  The data shown here have the background $d$ subtracted and envelope function $I(T)$ normalized away, i.e. they correspond to the pure beat signal $S_b(T)$.  The interference between the two beat frequencies, as described in the text, is evident.  For this data, $B = 540$ mG, corresponding to the average beat frequency $\bar{\omega}_b \cong 1.35$ MHz; the electric field amplitude $\mathcal{E} = 118$ V/cm; and we determine $\Delta \omega_b \cong 2 \pi \times 11.4$ kHz. \label{fig:beatbeat} }
\end{figure}

To obtain this data, we deliberately ran with a large magnetic field $B$ ($B \approx 0.5$ G) to enhance the difference $\Delta_{N_+} - \Delta_{N_-} \cong 2(g_{N_+}-g_{N_-})\mu_B B$.  From the fitting procedure described above we extract the pure beat signal $S_b(T)$, defined as the signal with the background $d$ subracted, the modulation envelope $I(T)$ normalized away, and the beat offset subtracted: $S_b(T) \equiv \frac{S(T) - d}{\alpha I(T)} - 1$.  Under these conditions $S_b(T)$ contains two Zeeman quantum beat signals of the same amplitude but nearly-equal frequencies; hence the term $c \cos{(\omega_b T + \phi)}$ in the usual expression for $S(T)$ [Eq. \ref{eq:fitequation}] must be replaced by $c[\cos{(\omega_{b+}T+\phi)} + \cos{(\omega_{b-}T+\phi)}]/2 = c \cos{(\bar{\omega}_{b}T+\phi)}\cos{(\Delta \omega_{b}T/2)}$.  Here $\bar{\omega}_b = (\omega_{b+}+\omega_{b-})/2$,  and $\Delta \omega_{b} = \omega_{b+}-\omega_{b-}$.  The resulting pure beat signal is $S_b(T) \propto e^{-T/T_b}[\cos{(\bar{\omega}_{b}T+\phi)}\cos{(\Delta \omega_{b}T/2)}]$. We determined $\Delta \omega_{b} = 2(g_{N_+}-g_{N_-})\mu_B B$ by finding the time $T_0$ of the first node in the envelope function of $S_b(T)$, such that $\cos{(\Delta \omega_{b}T_0/2)} = 0$. A signal of this type is shown in Fig. \ref{fig:beatbeat}.

%\subsection{theory with g(E)}

The modification of the $g$-factors of these states as a function of $\mathcal{E}$ is determined not only by mixing within the $\Omega$-doublet $J\! = \! 1$ levels, but also by mixing with the more distant $J\! = \! 2$ levels. 
%We define $g_H (g_L)$ to be the $g$-factor of the $\Omega$-doublet state with higher (lower) energy, so that e.g. $g_H(\mathcal{E} \! =\! 0) = g_{P-}$, while $g_H = g_{N_-}$ for sufficiently large $\mathcal{E}$. 
We calculate $g_{H,L}(\mathcal{E})$ as follows. We initially find eigenstates $\ket{\psi_{H,L}(M)}$ of the two-state subspace $\left( \ket{J=1,M,P=-1}, \ket{J=1,M,P=+1} \right)$ under the Hamiltonian $H_{St}^{(1,M)}$ (Eq. \ref{eq:2by2StarkHamiltonian}).
%\begin{equation}
%H_{St}^{(1)} = \left(
%\begin{array}{cc} %
%\Delta_{\Omega_1} & \frac{\mu_a \mathcal{E} M}{2} \\
%\frac{\mu_a \mathcal{E} M}{2}  & 0
%\end{array}
%\right).
%\end{equation}
%Interesting cases include e.g. $\mu_a \mathcal{E} \ll \Delta_{\Omega_1}$, in which case $\ket{\psi_{H}(M)} = \ket{J=1,M,P=-1} + \eta_P\ket{J=1,M,P=-1}$ (with $\eta_P(M) \approx -M \mu_a \mathcal{E}/\Delta_{\Omega_1}$); and $\mu_a \mathcal{E} \gg \Delta_{\Omega_1}$, where $\ket{\psi_{H}(M)} = \ket{J=1,M,N=-1} + \eta_{\mathcal{E}}(M)\ket{J=1,M,N=+1}$ (with $\eta_{\mathcal{E}}(M) \approx M \Delta_{\Omega_1}/(2\mu_a \mathcal{E})$).  
Next we include the effect of Stark-induced admixtures of the $J\! =\! 2$ state.  Since $\mu_a \mathcal{E} \ll B_r$ for all cases of interest, we use first-order perturbation theory to describe this mixing.  The perturbed eigenstate is then $\ket{\widetilde{\psi}_{H,L}(M)} =  \ket{\psi_{H,L}(M)} + \eta_{2+}(M)\ket{J=2,M,\Omega=+1} + \eta_{2-}(M)\ket{J=2,M,\Omega=-1}$, where $\eta_{2\pm}(M) \cong \bra{J'=2,M',\Omega'=\pm} H_{St}  \ket{\psi_{H,L}(M)}/(4B_r)$.  Calculation of matrix elements of the form $\bra{J'=2,M',\Omega'} H_{St} \ket{J=1,M,\Omega}$ entering this expression is outlined in Appendix B.   

The first-order Zeeman shift $\delta_{Z_{H,L}}$ of this perturbed state, due to the magnetic Hamiltonian $H_{mag}$, is given by $\delta_{Z_{H,L}} = \bra{\widetilde{\psi}_{H,L}(M)} H_{mag} \ket{\widetilde{\psi}_{H,L}(M)}$ and is calculated as follows.  Off-diagonal (in $J$) matrix elements of $H_{mag}$ are evaluated using the approximate form $H_{mag} \approx G_\parallel \mathbf{J}_e \cdot \mathbf{n} B = G_\parallel \Omega B$, where $G_\parallel \cong 1.86$ is the component of the molecular $G$-tensor along the internuclear axis.  Diagonal matrix elements of $H_{mag}$ are calculated in the $\ket{J,M,P}$ basis, where $\bra{J,M,P} H_{mag} \ket{J,M,P} = g_{P\pm}$.  This formulation is correct up to first order in both small factors $\Delta g(0)$ and $\mu_a \mathcal{E}/B_r$.  The $g$-factor difference $\Delta g(\mathcal{E}) = g_H(\mathcal{E}) - g_L(\mathcal{E})$ is of primary interest, since it determines the zero-order effectiveness of the $\Omega$-doublet comagnetometer.  The numerically calculated form of $\Delta g(\mathcal{E})$ is shown in Fig. \ref{fig:geff}, where the good agreement with our measured values is evident.
For the case $\mu_a \mathcal{E} \gg \Delta_{\Omega_1}$, a simple relation holds:
\begin{equation}
\Delta g(\mathcal{E}) \cong \frac{\Delta_{\Omega_1}}{\mu_a \mathcal{E}}~\Delta g(0) + \frac{3}{20}\frac{\mu_a \mathcal{E}}{B_r}~G_\parallel. \label{eq:Deltagformula}
\end{equation}

%\subsection{transition to how we really do experiment}

This method of simultaneously populating both states $\ket{\psi_{N_+}}$ and $\ket{\psi_{N_-}}$ could in principle be used for measurement of $d_e$.  However, it requires working at large values of both $B$ and $\mathcal{E}$ in order to clearly resolve $\Delta \omega_b$; this in turn increases the size of several systematic effects in the measurement of $d_e$.  Moreover, we found it difficult to attain shot-noise limited sensitivity to $\Delta \omega_b$ from this type of data (although a more sophisticated data analysis method might remove this problem).  For the moment, we have found it preferable to use a state preparation sequence that selectively populates either $N=+1$ or $N=-1$ states for a given laser pulse.  Two different double-resonance techniques were developed to achieve this goal, as described below.

\section{Oriented state preparation via Laser-RF double resonance}
%\subsection{idea of RF population}

The first double-resonance scheme for selectively populating one oriented molecular state worked as follows.  Throughout the excitation and detection sequence $\mathcal{E}$ was present.  A $\hat{z}$-polarized laser pulse populated the $\ket{J=1,M=0,P=-1}$ eigenstate.  Next, a short pulse of radiofrequency magnetic field $\hbox{\boldmath{$\beta$}}_{RF} = \beta_0 \cos{(\omega_{RF}T)} \hat{y}$ was applied to drive $M\! =\! 0 \rightarrow M'\! =\! \pm 1$ magnetic dipole (M1) transitions to the oriented molecular states.  When $\omega_{RF} = \omega_{RF_H} \equiv \delta_{St} ~ (\omega_{RF_L} \equiv \delta_{St} + \Delta_{\Omega_1})$, the higher (lower) energy state $\ket{\psi_{N_-}(0)} (\ket{\psi_{N_+}(0)})$ can be populated.  Successful population of a $\ket{\psi_{N\pm}}$ state will result in appearance of quantum beats in the fluorescence signal.  Note that in the limit of full mixing/polarization by $\mathcal{E}$, the M1 transition strengths to the two oriented states are identical. The RF pulse durations must be sufficiently short to ensure that the linewidth of the pulses is broad enough to coherently populate both $M=\pm 1$ sublevels, as well as to cover the inhomogeneous broadening in the Stark-shifted average transition frequency due to inhomogeneity in the DC electric field $\mathcal{E}$.  This in turn requires that the RF field strength $\beta_0$ be sufficiently strong to drive a $\pi$-pulse within this short time.

\begin{figure}
\includegraphics{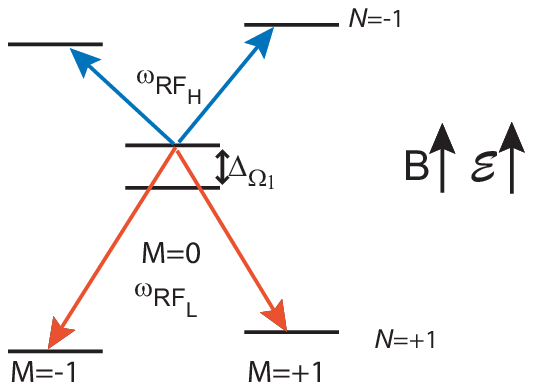}%
\caption{(color online) Schematic of the laser-RF double resonance method for population of oriented molecular states.  The laser pulse populates the $\ket{J\! =\! 1, M\! =\! 0, P\! =\! -1}$ state.  Then, a short RF pulse is tuned to drive the population to either $N=-1$ (blue with frequency $\omega_{RF_H}$) or $N=+1$ (red with frequency $\omega_{RF_L}$). \label{fig:RF}}
\end{figure}

%\subsection{RF apparatus}

To apply the RF magnetic field, square single-turn Helmholtz coils were constructed around the cell body with their axis along the laser beam, in the $\hat{y}$ direction.  The coils were connected in parallel to reduce their inductance and driven via coaxial feedthroughs.  Outside the vacuum chamber, it was necessary to impedance match the RF coils at both $\omega_{RF_H}$ and $\omega_{RF_L}$.  This was achieved using two separate matching networks, one for each frequency, which were switched in or out of the line using relays.
%These networks consisted of a common coaxial cable (with length $L$ chosen to be a half-integer multiple of the RF wavelength for both frequencies), plus a serial capacitor $C_1$ and a parallel capacitor $C_2$.  Typical values were $L \sim 10$ m, $C_1=8$ pF and $C_2$=25 pF for $\omega_{RF_H}$=33.6 MHz; and $C_1=15$ pF, and $C_2$=60 pF for $\omega_{RF_L}$=44.8 MHz.
The coils were driven by a broadband pulse amplifier (1 kW maximum output power) excited with an arbitrary waveform generator for pulse frequency and envelope control.  The RF field strength and homogeneity were measured using a calibrated miniature pickup coil, in a room-temperature mockup of the actual cell and heater geometry that reproduced all nearby conductive parts such as electrodes and heater elements.  RF field strength as large as $\beta_0 \sim 1-1.5$ G, corresponding to $\pi$-pulse times $\tau_{RF} < 1.1~\mu$s, could be applied at several different drive frequencies spaced by $\Delta_{\Omega_1}$.  The RF field was measured to be homogeneous to better than 20\% over the active volume of the cell, consistent with extensive numerical modeling of the system.

\begin{figure}
\includegraphics[width=3.3in]{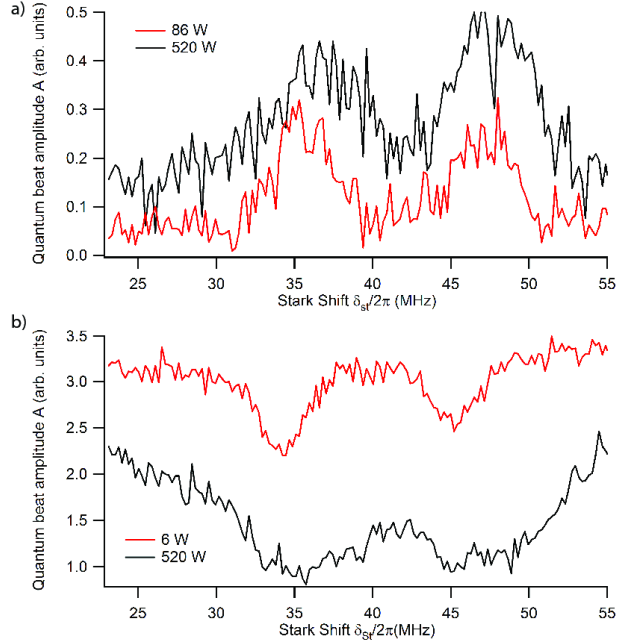}%
\caption{Quantum beat signals with the laser-RF double resonance scheme.  The quantum beat amplitude $A$ (defined as $A \equiv c \alpha$) is plotted as a function of Stark shift $\delta_{St}$ (a) Appearance of quantum beats following RF population of oriented molecular states.  The black (red) trace is at high (low) RF power, as indicated in the legend.  Even at the highest powers, the corresponding contrast is only about 10-20\% of that seen with direct laser state preparation (corresponding to $A \sim \! 3$ on this plot).  The lines broaden less than expected as the power is increased.  (b) Disappearance of quantum beats following RF depletion of laser-excited states in the presence of $\mathcal{E}$.  Beat depletion as large as 70\% was observed, and the width of these resonances was consistent with expectations due to RF power broadening.  \label{fig:b_app_diss}}
\end{figure}

%\subsection{RF results}
Typical results for RF excitation are shown in Fig. \ref{fig:b_app_diss}.  Here, $\omega_{RF}$ was fixed at either $2 \pi \times 33.6$ MHz or $2 \pi \times 44.8$ MHz and the electric field strength $\mathcal{E}$ was scanned so that $\ket{\psi_{N_-}}$ and $\ket{\psi_{N_+}}$ came successively into resonance.  Clear peaks in the quantum beat signal were observed at the expected positions.  However, the maximum contrast achieved was only about 10-20\% of that seen with direct laser state preparation as described previously.  This poor RF excitation efficiency was inconsistent with expectations based on the measured RF field strength and homogeneity.  Moreover, the width of the resonance peaks exhibited no sign of power broadening, which should have been marked at high RF power.

In order to investigate this effect further, we studied the inverse process of driving $M\! = \! \pm 1 \! \rightarrow \! M\! =\! 0$ transitions.  Here, the states $\ket{\psi_{N_+}}$ and $\ket{\psi_{N_-}}$ were initially populated using an $\hat{x}$-polarized laser as in the previous section.  In this case, driving the RF transition will result in disappearance of the quantum beats.  (Note that for each of the two resonant conditions, each of the two initial states is resonant with a different final state $\ket{J\! = \! 1, M\! = \! 0, P\! = \! \pm 1}$, so complete disappearance of the beat contrast should be possible.)  As shown in Fig. \ref{fig:b_app_diss}, this effect was indeed observed, but showed substantially greater efficiency (approaching $\sim\! 70\%$) than in the beat appearance data (where the $M\! =\! 0 \rightarrow M'\! =\! \pm 1$ transitions needed for preparation of oriented molecular states were driven).  Moreover, the width of the disappearance resonances was consistent with the expected level of power broadening.

The reason for the discrepancy between the apparent efficiency of excitation and de-excitation in this system has not been fully understood.  It appears to indicate that at the high drive fields expected to be necessary for a $\pi$-pulse condition, the Zeeman coherence between the $|M| = \pm 1$ levels actually is being dephased more rapidly than the Rabi oscillation period.  We suspect that this effect might arise from the mixed-parity nature of the $|M|\! =\! \pm 1$ states, as follows.  The RF magnetic field $\hbox{\boldmath{$\beta$}}_{RF}$ induces an RF electric field $\hbox{\boldmath{$\epsilon$}}_{RF}$.  Boundary conditions from the in-cell electrodes cause $\hbox{\boldmath{$\epsilon$}}_{RF} \parallel \hat{z}$ over the active volume.  This field has a node in the center of the cell, but at the edge of the electrodes (a radial distance $d = 1"$ from the center) reaches the substantial value $\epsilon_{RF}(R=d) \sim \beta_{RF} (d \omega_{RF}/c) \sim 5$ V/cm.  This in turn induces a Stark modulation of the resonance energies $\omega_{RF_{H,L}}$, at the drive frequency $\omega_{RF}$.  The depth of this modulation can be substantial, e.g. $\epsilon_{RF}(R=d) \mu_a/2 \sim 4$ MHz, and is inhomogeneous at the 100\% level over volume of the cell.  It seems plausible that this field could lead to the observed dephasing, although no specific model we have applied appears to explain our results.

%\subsection{reason RF failed}

As a result of the inefficient and poorly-understood results from RF preparation of the oriented molecular states, we investigated an alternate population scheme. This method, described below, uses microwave Raman transitions through the intermediate $J=2$ rotational state.  The much stronger electric dipole amplitudes associated with the rotational transition yield substantially greater excitation efficiency, and the microwave method has also proven easier to implement than the RF technique.

\section{Oriented state Preparation via laser-microwave Raman double resonance}
\label{sec:MSP}

\begin{figure}
\includegraphics[width=5.0in]{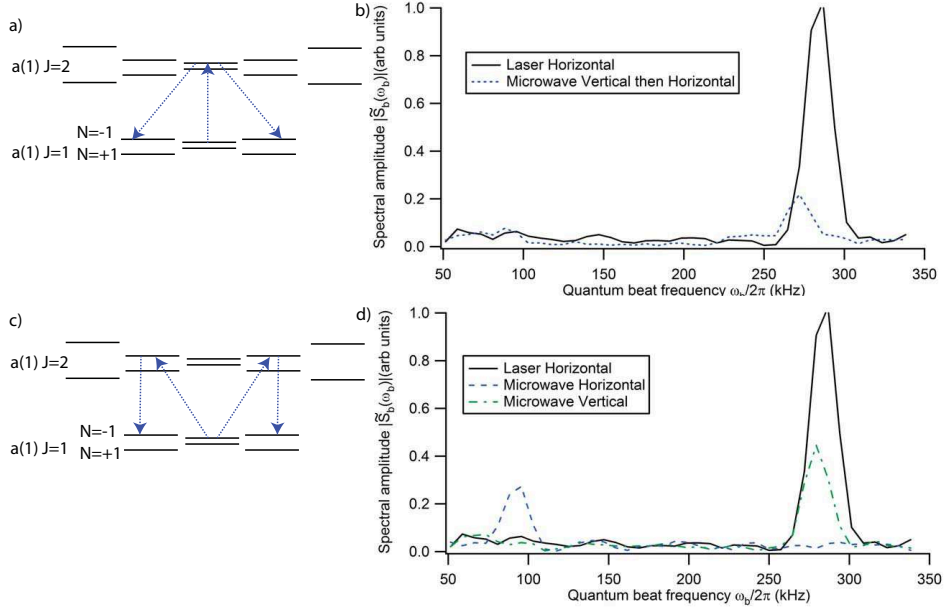}
\caption{(color online) Laser-microwave Raman double resonance state preparation.  The plots show data from selective preparation of oriented states using microwave adiabatic passage.  The r.m.s. amplitude spectrum $|\tilde{S}_b(\omega_b)|$ of the pure beat signal $S_b(T)$ (see Fig. \ref{fig:beatbeat}) is plotted as a function of the beat frequency $\omega_b$.  In both plots, the black solid traces show beat data with $\hat{x}$ laser polarization preparing an incoherent superposition of oriented states, as in Sec. VI.  This provides a baseline for comparison to the case where oriented states are prepared.  Preparation of oriented states began with $\hat{z}$ laser polarization, preparing the $\ket{J\! =\! 1,M\! =\! 0,P\! =\! -1}$ state.  Oriented states were populated with microwave excitation from this initial state. (a) Energy level diagram for the xz pulse sequence. (b) Results from the zx pulse sequence.  The dashed blue trace shows the beat spectrum after both pulses, showing $\sim\! 25\%$ transfer efficiency.  (No beats were observed after the first pulse.) For this data, the first (second) pulse had a frequency sweep centered around $\omega_{\mu 1} = 2 \pi \times 28.220$ GHz ($\omega_{\mu 2} = 2 \pi \times 28.193$ GHz). (c) Energy level diagram for the zx pulse sequence. (d) Results from the xz pulse sequence.  The dashed blue (dot-dashed green) trace shows the spectrum after the first (second) pulse.  After the first pulse, beats appear at the frequency corresponding to the Zeeman splitting between $\ket{J\! =\! 2, M\! = \pm 1}$ states.  After the second pulse, $\sim \! 45\%$ transfer efficiency is achieved.  The first (second) pulse was centered around $\omega_{\mu 1} = 2 \pi \times 28.217$ GHz, ($\omega_{\mu 2} = 2 \pi \times 28.190$ GHz).   For both plots, $\mathcal{E} \cong 39$ V/cm. Each microwave pulse had 2 $\mu$s duration, during which the frequency was swept linearly over a range $d\omega = 2 \pi\times 11$ MHz.  The second pulse arrived $0.5 \mu$s after the first ended. Both pulses had nominal power of $\sim \! 1$ W.
\label{fig:microwaves}}
\end{figure}

%\subsection{idea of MSP}
As in the RF method, a $\hat{z}$-polarized laser pulse populated the $\ket{J=1,M=0,P=-1}$ eigenstate in the presence of $\mathcal{E}$.  Next, one resonant microwave pulse with polarization $\hbox{\boldmath{$\epsilon$}}_{\mu 1}$ and frequency $\omega_{\mu 1}$ excited these molecules to sublevels of the $J=2$ manifold of states.  Then, a second pulse with orthogonal polarization $\hbox{\boldmath{$\epsilon$}}_{\mu 2}$ and frequency $\omega_{\mu 2} = \omega_{\mu 1} - \delta_{St}$ or $\omega_{\mu 2} = \omega_{\mu 1} + \delta_{St} + \Delta_{\Omega_1}$ stimulated deexcitation to the $\ket{\psi_{N_+}}$ or $\ket{\psi_{N_-}}$ state, respectively (see Fig. \ref{fig:microwaves}). The linewidth $\Gamma_{\mu \mathrm{w}}$ of these microwave transitions is determined by Doppler broadening and/or inhomogeneity in $\delta_{St}$; under all conditions $\Gamma_{\mu \mathrm{w}} \ll \Delta_{\Omega_1}, \delta_{St}$ so that the microwave pulses can easily resolve the states with well-defined molecular orientation.  As in the RF scheme, the pulse durations must be sufficiently short to ensure coherent population of the $M\! =\! \pm 1$ states.

%\begin{figure}
%\includegraphics{pop}
%\caption{Laser-microwave Raman double resonance method for preparation of oriented molecular states. Here the zx %sequence (see text) for preparing a state with $N = -1$ is shown. The laser populates the $\ket{J\!=\!1,M\!=\!0,P\! %= \! -1}$ state.  Next, a $\hat{z}$-polarized microwave pulse is used to transfer the population to the state %$\ket{J\!=\!2, M\!=\!0, P\! = \! +1}$.  Finally, the population is transferred to the coherent superposition state %$\ket{\psi_{N_-}}$, with a $\hat{x}$-polarized microwave pulse.  The state $\ket{\psi_{N_+}}$ can be accessed in the %same way, simply by tuning the frequency of the $\hat{x}$-polarized microwave pulse.}
%\label{fig:population}
%\end{figure}

We investigated these microwave Raman transitions both with $\hbox{\boldmath{$\epsilon$}}_{\mu 1} = \hat{x}, \hbox{\boldmath{$\epsilon$}}_{\mu 2} = \hat{z}$ (xz sequence); and with $\hbox{\boldmath{$\epsilon$}}_{\mu 1} = \hat{z}, \hbox{\boldmath{$\epsilon$}}_{\mu 2} = \hat{x}$ (zx sequence).  These sequences have different advantages and disadvantages.  In the xz sequence, the first step creates an intermediate state $\ket{\psi_{i(xz)}} \propto \ket{J\! =\! 2, M\! =\! 1} - \ket{J\! =\! 2, M\! =\! -1}$.  After the second step, any residual population in this $J=2$ superposition state can produce quantum beats that can complicate the data analysis (although these beats are at a different frequency than the desired $J=1$ beats).  Furthermore, Stark mixing in the $\ket{J\! =\! 2, |M|\! =\! 1}$ states is incomplete over much of the useful range of $\mathcal{E}$ (both because $\Delta_{\Omega_2} = 3 \Delta_{\Omega_1}$, and because the Stark matrix elements are smaller between $J=2$ sublevels).  This means that the transition amplitudes in each step vary as $\mathcal{E}$ changes, again complicating data analysis.  The zx sequence (where the intermediate state is always $\ket{\psi_{i(zx)}} = \ket{J\! =\! 2, M\! =\! 0, P\! =\! +1}$) does not suffer from these difficulties.  However, the second step of this sequence has a particularly small transition amplitude (due to unfavorable Clebsh-Gordan coefficients), making it difficult to saturate fully.

Microwave generation starts with a two channel arbitrary waveform generator which outputs waveforms at a frequency of $\sim$32 MHz.  Each channel determines the excitation frequency for one step of the Raman process.  At various points in our study of this process, we investigated $\pi$-pulse excitation (by varying the area of the pulse envelope), as well as use of adiabatic passage (using linear frequency chirps).  The $\sim$32 MHz signal is multiplied by 32 times to $\sim$1 GHz, filtered, and mixed with the 13.1 GHz signal from a phase locked
dielectric resonator oscillator.  A waveguide filter removes the low sideband leaving a signal at 14.1 GHz, which is doubled to match the splitting between the $J=1$ and $J=2$ states (nominally $\omega_\mu \approx 2 \pi \times$28.2 GHz) and amplified (up to 4 W output).  A fast switch is used to direct the 14.1 GHz signal through either of two doubler/amplifier setups, which feed respectively into the $\hat{x}$- or the $\hat{z}$-polarized mode of an orthomode transducer whose output is connected to a conical horn.

The horn output is located inside a 2" inner diam. Teflon tube, which slides over one of the 2" diam. quartz rods that extends into the vacuum chamber (ending 1 cm from the vapor cell). The Teflon tube and quartz rod act as a multimode waveguide, minimizing the angular spread of the microwave beam before it reaches the vapor cell.  The microwave beam propagates opposite the laser beam, in the $-\hat{y}$ direction.  The microwaves spread rapidly after exiting the light pipe, and undergo significant reflections from the sapphire cell windows; the power exiting the light pipe on the opposite side of the cell is reduced from the input power by $\sim 10$ dB.  In addition, the amplitude of $\hat{x}$-polarized microwaves must vanish on the cell electrodes (parallel to the $x\! -\! y$ plane) due to the usual boundary conditions.  Hence, there is significant inhomogeneity in the microwave field amplitude over the active cell volume.  Under most conditions the power delivered to the center of the cell was much greater than needed to saturate the transition; use of adiabatic passage rather than $\pi$-pulses typically yielded somewhat higher excitation efficiency.

The efficiency for driving microwave transitions was monitored by the appearance and/or disappearance of quantum beats in fluorescence, as discussed in section VII.  It was possible and often helpful to track quantum beats at the two different frequencies associated with the $J=1$ and $J=2$ states, by observing the power spectrum of the data.  As mentioned above, the transition matrix elements varied considerably depending on the microwave polarization, the value of $\mathcal{E}$, and the specific $J,M$ sublevels being addressed.  Roughly speaking, however, at the maximum available power optimized $\pi$-pulse transfer was achieved with pulse durations of $\sim \! 100-300$ ns.  Optimal adiabatic passage pulses were $\sim\! 0.5-2 ~\mu$s in duration, with a total frequency sweep of $\sim \! 2-30$ MHz (symmetric about the resonant frequency).  Under a variety of conditions we have achieved $\gtrsim 50\%$ transfer efficiency in single microwave pulses, and $\gtrsim 30\%$ for the Raman transfer from $\ket{J=1,M=0,P=-1}$ to $\ket{\psi_{N_+}}$ or $\ket{\psi_{N_-}}$. Typical data is shown in Fig. \ref{fig:microwaves}.

%$\hat{z}$ polarized microwave pulse adiabatically transfers the
%population to the $\ket{a(1),J\!=\!2,M\!=\!0,e}$ level as shown in Fig. \ref{fig:population}.  This pulse is short %($\sim$100 ns) with large modulation (20 MHz).  A second longer $\hat{x}$ polarized
%pulse ($\sim$1 $\mu$s) with smaller deviation (6 MHz) is used to
%transfer the population to the $\ket{J\!=\!1,M\!=\!+1}\!+\!\ket{J\!=\!1,M\!=\!-1}$ state.

%\subsection{apparatus for MSP}

We note in passing that the lineshape for (one-step, single-frequency) microwave excitation depends on the homogeneity of $\mathcal{E}$ in the detection region, as a result of the different Stark shifts of the initial and final states.  This effect made it possible to determine the r.m.s. variation $\delta\mathcal{E}/\mathcal{E} = 1\%$, in good agreement with expectations based on the distribution of $\mathcal{E}$ found in finite element calculations with our electrode geometry.

\section{Oriented state preparation: laser excitation and microwave erasure}

\begin{figure}
\includegraphics[width=3.3in]{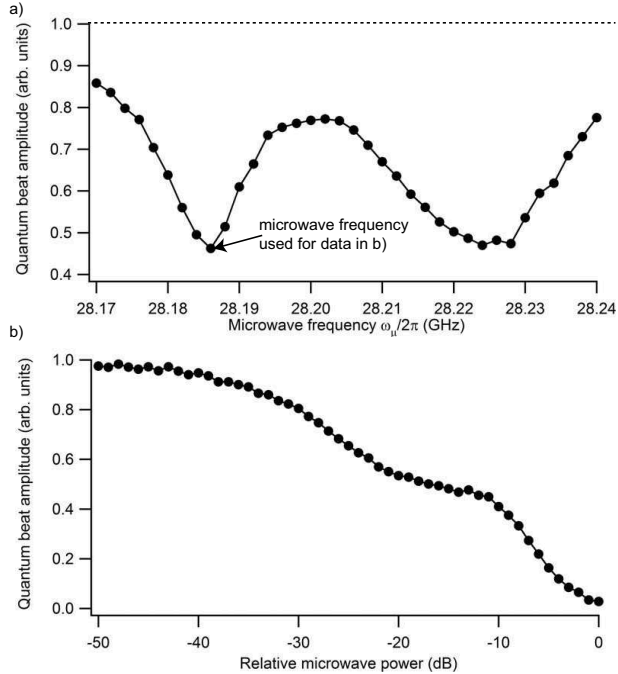}
\caption{Confirmation of the microwave ``erasure'' method.  (a) Plot of quantum beat amplitude $A = c \alpha$ versus microwave frequency, showing two dips at the expected resonant frequencies.  The dashed horizontal line corresponds to the full beat amplitude seen with no microwaves applied.  At each resonance, the beat contrast drops to approximately half its original size, indicating that the Zeeman coherence in one state has been eliminated.  For these data, 0 dB microwave power corresponds to $\sim 1$ W transmitted through the vapor cell, and the pulse duration was $= 9 \mu$s.  (b) Plot of beat amplitude $A$ versus microwave attenuation, at the (resonant) microwave frequency indicated by the arrow in (a).  Initially increasing the power increases the efficiency of "erasure" until the beats from one pair of states are completely erased and $A$ has dropped by one half.  Further power broadens the lines further, until the 2nd pair of states is also ``erased''.  Nearly-pure oriented states are prepared at the plateau indicating $\sim\! 50\%$ ``erasure'' in the range of powers $-15 \leftrightarrow -10$ dB. 
\label{fig:mweras}}
\end{figure}

To date, the most efficient and robust method for creating states with simultaneous molecular orientation and spin alignment uses the technique we refer to as ``microwave erasure''.  This method begins as described in section VI: $\hat{x}$-polarized laser excitation is used to prepare an equal-weight, incoherent mixture of $\ket{\psi_{N_+}}$ and $\ket{\psi_{N_-}}$.  Next, a microwave pulse, with polarization $\hbox{\boldmath{$\epsilon$}}_{\mu} = \hat{x}$, is applied for a long duration $\tau_{erase} \gg \delta_Z^{-1}$ so that the spin alignment precesses over many complete cycles during the pulse.  The microwave frequency is resonant with either the $\ket{\psi_{N_+}}\rightarrow \ket{J=2,|M|=2,N\! =\! +1}$ or the $\ket{\psi_{N_-}}\rightarrow \ket{J=2,|M|=2,N\! =\! -1}$ transition.  The microwave power is chosen to be moderately high, so as to induce a transition Rabi frequency $\Omega_{R\mu}$ such that $\delta_Z \ll \Omega_{R\mu} \lesssim \Delta_{\Omega_1}$.  The combination of long pulse duration, large transition Rabi frequency, and microwave field inhomogeneity over the cell can lead to essentially complete dephasing of quantum beats from the $J\! =\! 1$ state that is in resonance with the pulse.  (This ``erasure'' is similar to the behavior observed in the beat disappearance data for the RF transitions; however, here a single oriented state can be spectroscopically addressed as a result of the different $\Omega$-doublet and Stark splittings in $J=2$ vs. $J=1$ states.)  The beats from the remaining oriented level thus have $50 \%$ of the maximum possible contrast, superior to any of the other methods tried.  The complete dephasing of the other state was verified in a variety of ways.  For example, at high values of both $\mathcal{E}$ and $B$ where the quantum beats of the two $\ket{\psi_{N\pm}}$ levels can be distinguished (as in section VI), the ``erasure'' of one beat frequency eliminates the interference between the beats.  At lower values of $\mathcal{E}$ and $B$, the contrast is monitored as a function of microwave frequency and power.  Near resonance and over a substantial range of microwave pulse power and duration, the contrast after ``erasure'' remains at 50\% of its value when the microwave pulse is not applied (see Fig. \ref{fig:mweras}).  This indicates that a single oriented state has been selectively dephased.

\section{Limitations of current method for detecting spin alignment precession}

Next, we turn to alternate methods for detecting the precessing spin alignment in the oriented molecular states.  The method of quantum beats in fluorescence has a number of features that induce serious limitations in the detection efficiency of useful signals, and in addition give rise to large background signals.  The effect of both factors on the ultimate uncertainty in measurement of $d_e$ was noted in Section IV.  In order to understand the motivation for the new techniques discussed here, it is useful to review the specific problems associated with the present technique.

The overall efficiency of fluorescence detection is determined by several factors.  The values of these factors varied over the course of the data described in this paper.  Here we report the best (and most current) values: the solid angle of collection into the light pipes $\eta_\Omega \approx 2\times 0.04$ (with two light pipes); the efficiency of transmission through the optical filters $\eta_f \approx 0.1$; the branching ratio for the detected vibrational decay channel $\eta_{BR} \approx 0.5$; and the photomultiplier quantum efficiency $\eta_{q.e.} \approx 0.15$.  Both $\eta_\Omega$ and $\eta_f$ were determined by Monte-Carlo simulations and also checked by direct measurements.  The total detection efficiency is thus $\eta_{tot} = \eta_\Omega \eta_f \eta_{BR} \eta_{q.e.} \approx 5 \times 10^{-4}$.

The backgrounds in the current detection scheme are closely linked to the detection method and its efficiency.  Since the a(1) state decays primarily to the X state, the detected fluorescence must be close in wavelength to that of the excitation laser.  Detection at the laser wavelength leads to such large levels of scattered light that the PMTs (despite use of a proven gating circuit \cite{Yoshida1989}) cannot recover before the a(1) state has decayed. Hence we must use interference filters to select decays from the a(1) state at different wavelengths, corresponding to different vibrational final levels in the X state.  Because the transmission of interference filters shifts towards the blue at non-normal incidence, scattered light remains a problem unless the selected fluorescence wavelength is to the blue of the laser.  For this reason, we excite the X$(v=1) \rightarrow$ a$(v'=5)$ transition at $\lambda = 570$ nm and detect the a$(v'=5) \rightarrow$ X$(v=0)$ transition at $\lambda = 548$ nm using narrowband interference filters.  (Note that at our cell temperature $T \sim 700^\circ$ C, the X$(v=1)$ state thermal population is roughly $1/3$ that in the X$(v=0)$ state, thus reducing the excitation efficiency from its optimal value by the same factor.)  The filter bandwidth (FWHM = 16 nm) is chosen as a compromise between wide-angle transmission of the fluorescence light (which improves with greater bandwidth) vs. increased levels of scattered light and blackbody radiation.  Finally, the inability to resolve different rotational levels in the decay fluorescence leads to severely reduced beat contrast.

\section{Spin Alignment Detection via Optical Double Resonance}
%\subsection{idea of reexcitation}

\begin{figure}
\includegraphics[width=3.3in]{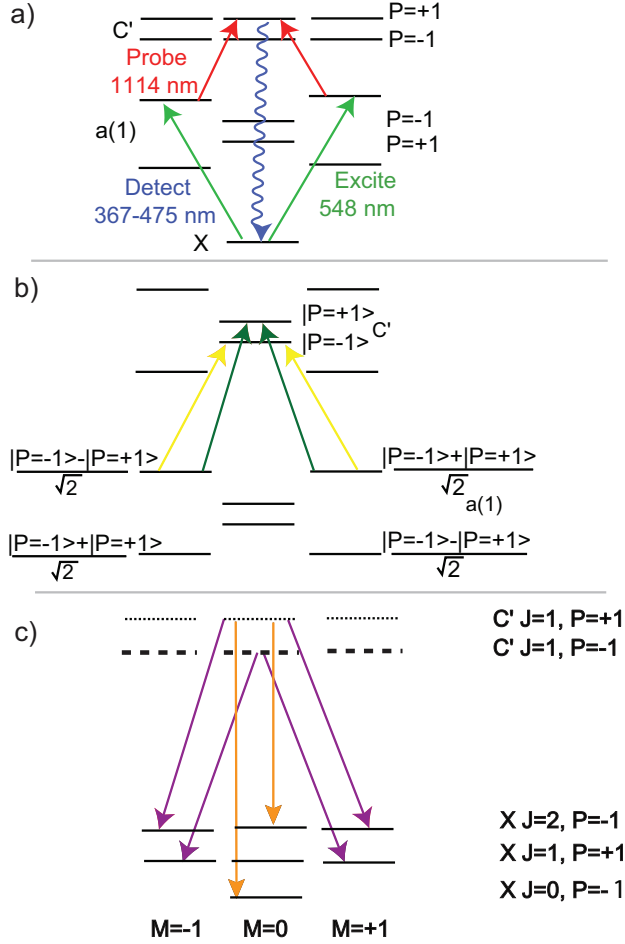}%
\caption{(color online) Optical double resonance probe schemes using the a-$\mathrm{C}'$ transition.  (a) General features of the scheme, illustrated for the case of  $\mathcal{E}=0$.  Here both excitation and probe lasers have $\hat{x}$ polarization, and a single sublevel of the $\mathrm{C}'$ state is populated.  (b) Probe excitation of an oriented molecular state.  Both $P\! = \!+1$ and $P\! = \! -1$ sublevels of the $\mathrm{C}'$ state are coherently populated by the probe laser, with their relative amplitude oscillating in time as the spin alignment of the $\ket{\mathrm{a}, J=1}$ state precesses. (c) Decays from the $\ket{\mathrm{C}', J\! =\! 1,M\! =\! 0,P\! =\! \pm 1}$ superposition state created by probing oriented molecular states.  Population oscillates between the $P\! =\! +1$ (dots) and $P\! =\! -1$ (dashes) states.  By collecting only $\hat{z}$-polarized fluorescence (orange arrows), only the $P=+1$ population is detected and quantum beats with high contrast ($c \cong 1$) should be observed.
\label{fig:fourpart}}
\end{figure}

These considerations led us to consider a new detection scheme.  The basic idea, shown in Fig. \ref{fig:fourpart}, is to probe the a(1) state spin alignment with a laser tuned to the a$(v=5)\rightarrow$$\mathrm{C}'(v'=5)$ transition (at wavelength $\lambda _{\mathrm{aC}'} \approx 1114$ nm)~\cite{firstPbO}, and subsequently detect the resulting $\mathrm{C}'$(1) state population via its rapid decay fluorescence to the X state.  This scheme has several advantages.  The beat contrast $c$ is expected to increase up to nearly 100\%, since the double resonance method probes only the desired $\ket{\mathrm{a},J\! =\! 1, M\! =\! \pm 1}$ superposition state.  In addition, the $\mathrm{C}'$-X fluorescence occurs at blue wavelengths (see below), which leads to three improvements.  First, the PMT quantum efficiency can be 3-4 times larger than at present; second, blackbody radiation is dramatically smaller; and third, scattered light from both excitation lasers should be trivial to eliminate, so that the optical filter angular acceptance can be increased.  The advantages are offset by a few difficulties. The most serious of these is related to the small cross section for the a$\rightarrow \mathrm{C}'$ transition~\cite{firstPbO}: very large laser power (see below) would be needed to saturate the a$\rightarrow$$\mathrm{C}'$ transition over the vapor cell.  In addition, the finite lifetime of the $\mathrm{C}'$ state ($\tau_{\mathrm{C}'} \approx 3.4 ~\mu$s) acts as a low-pass filter to diminish the contrast of beats with frequency $\omega_b \gtrsim \tau_{\mathrm{C}'}^{-1}$.  Nevertheless, we project that the overall gain in sensitivity to $d_e$ can be substantial using this detection method, and we thus have investigated it in some detail.

%\begin{figure}
%\includegraphics{Ceng}%
%\caption{(color online) Optical double resonance scheme for %detecting spin alignment precession.  The a(1) $J=1$ state %is populated as described in section \ref{sec:MSP}.  An %$\hat{x}$-polarized probe laser transfers information on the %relative phase between the $M=\pm 1$ sublevels of the %a(1)$(J=1)$ states into the population of the %\mathrm{C}'$$(M=0)$ state.  The $\mathrm{C}'$ state %population is monitored by detecting fluorescence along the %$\mathrm{C}'\rightarrow$X transition.\label{fig:Ceng}}
%\end{figure}

%\subsection{results with no E field}

Initial tests of this new detection scheme were conducted as follows. The state $\ket{\psi_{P-}^{(0)}}$ was populated as in section III, using an $\hat{x}$-polarized laser with $\mathcal{E} = 0$.  After a delay time $T$, the ensuing state $\ket{\psi_{P-}(T)}$ was probed by a second laser pulse, tuned to the Q1 line ($J\! =\! 1 \rightarrow J'\! =\! 1$) of the a(1)$[v=5]\rightarrow$$\mathrm{C}'$(1)$[v'=5]$ transition and with polarization $\hbox{\boldmath{$\epsilon$}}_{probe}=\hat{x}$.  This results in population of the $\ket{\mathrm{C}',J\! =\! 1,M\! =\! 0,P\! =\! +1}$ state, with probability $P_{\mathrm{C}'}$ determined by the transition matrix element:
\begin{eqnarray}
P_{\mathrm{C}'} & \propto & |\bra{\mathrm{C}',J\! =\! 1,M\! =\! 0,P\! =\! +1}\hbox{\boldmath{$\epsilon$}}_{probe}\cdot\mathbf{r}\ket{\psi_{P-}(T)}|^2 \nonumber \\
& = & |(e^{-i\delta_Z T}\bra{\mathrm{C}',J\! =\! 1,M\! =\! 0,P\! =\! +1} x \ket{J\! =\! 1,M\! =\! +1,P\! =\! -1} \nonumber\\
& \; & + e^{i\delta_Z T}\bra{\mathrm{C}',J\! =\! 1,M\! =\! 0,P\! =\! +1} x \ket{J\! =\! 1,M\! =\! -1,P\! =\! -1})|^2/2 \nonumber\\
& \propto & 1+ \cos{(2\delta_Z T)}.
\end{eqnarray}
Hence, the population of the $\mathrm{C}'$ state reflects the direction of spin alignment in a(1) state (see Fig. \ref{fig:fourpart}). This $\mathrm{C}'$ state population can be detected by monitoring its decay fluorescence, yielding a signal $S(T) \propto 1+ \cos{(2\delta_Z T)}$ with 100\% contrast ($c=1$).

%\begin{figure}
%\includegraphics{Cstruct}%
%\caption{(color online) Optical double resonance probe scheme with $\mathcal{E}=0$.  a)The a(1) state is excited %with $\hat{x}$ polarized
%light, preparing a superposition of the $\ket{J\! =\! 1,M\! =\! \pm 1,P\! =\! -1}$ levels.  The spin alignment of %this state is probed by
%reexciting from this state to the $\ket{J'\! =\! 1,M'\! =\! 0,P'\! =\! +1}$ sublevel of the $\mathrm{C}'$ state, %and monitoring its population via its decay fluorescence to the ground state.  Quantum beats with high contrast ($c %\cong 1$) are expected.
%end{figure}

%\subsection{experimental details}
For these measurements, the 1114 nm light was generated by a second Nd:YAG-pumped pulsed dye laser, downshifted to the infrared with a gaseous H$_2$ Raman converter cell (Light Age).  The pulses had duration of $\sim \! 7$ ns, bandwidth (averaged over many pulses) of $\sim 5$ GHz, and power of $\sim 0.7$ mJ/pulse.  To detect the $\mathrm{C}'$-X fluorescence, a broad color glass filter replaced the usual interference filter in front of a PMT, allowing for the detection of the decays from $\mathrm{C}'(v\!=\!5)$ to X($v'\!=\!0\!-\!9$), corresponding to wavelengths $\lambda_{\mathrm{C}'\mathrm{X}} \approx 370-475$ nm.  The variable time delay $T$ was controlled with the computer used for data acquisition.

\begin{figure}
\includegraphics[width=3.31in]{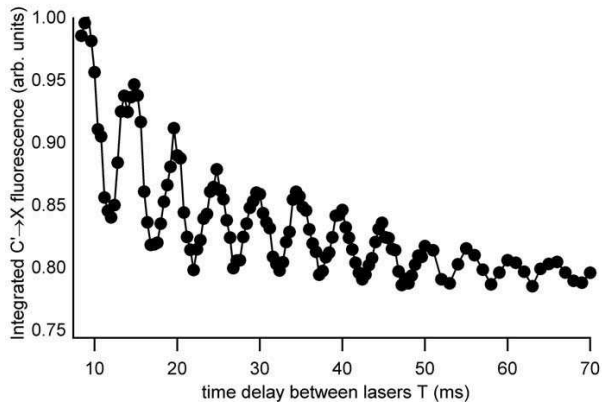}%
\caption{Observed quantum beats using optical double resonance probe detection, with no applied electric field.  The $\mathrm{C}'$$\rightarrow$X decay fluorescence signal is plotted as a function of time delay between the $\hat{x}$ polarized excitation laser [driving the X$\rightarrow$a(1) transition] and the $\hat{x}$ polarized probe laser [driving the a(1)$\rightarrow$$\mathrm{C}'$ transition]. As expected, near 100\% contrast is observed.  The DC background level in this data is believed due to light leakage through or around the low-quality blue-transmitting color glass filter used here.  Subsequent measurements indicated that negligible blackbody radiation is present over the range of wavelengths corresponding to $\mathrm{C}'$$\rightarrow$X fluorescence. \label{fig:Cbeats}}
\end{figure}

Typical data from these experiments is shown in Fig. \ref{fig:Cbeats}.  Here, as expected the signal has (nearly) 100\% contrast, validating the basic concept of this detection scheme.  From measurements of absolute signal sizes we can deduce the strength of the a-$\mathrm{C}'$ transition, and find it to be in reasonable agreement with estimates based on our previous measurement of the $\mathrm{C}'\rightarrow$a decay rate~\cite{firstPbO}.  This enables calculation of the saturation intensity for this transition (see below).  We note in passing that spectroscopy of neighboring rotational lines in the a-$\mathrm{C}'$ band allowed us to determine the $\mathrm{C}'$$(v=5)$ rotational constant $B_{v=5} = 0.2395(3)$ cm$^{-1}$ and vibrational energy $G(v=5) = 2625.1(1)$ cm$^{-1}$ (taking $T_e = 24947$ cm$^{-1}$ \cite{Herzbergconstants}).  We also determined an upper bound on the $\Omega$-doubling parameter $q_{\mathrm{C}'(v=5)} < 60$ MHz. These are consistent with, but more accurate than previous determinations \cite{Herzbergconstants}.

%\subsection{theory of no E field}

%\begin{figure}
%\includegraphics[width=3.3in]{noEdecay}%
%\caption{(color online) With $\mathcal{E}=0$ the decays $C'\rightarrow$X transition is expected only along a Q %branch. Here high contrast quantum beats are expected to be observed.\label{fig:noEdecay}}
%\end{figure}

Next we discuss the application of this method to detection of states with simultaneous molecular orientation and spin alignment.  Assume that such a state has been prepared, e.g. by any of the methods described earlier.  Consider the signal resulting from probing the spin alignment of the state by excitation to the $\mathrm{C}'$ state as just described.  Because of the parity mixing in the oriented molecular state $\ket{\psi_{N\pm}}$, and because the $\Omega$-doublet splitting in the $\mathrm{C}'$ state is too small to resolve with the laser, the probe laser will excite transitions to both (unmixed, definite parity) $M=0$ states in $\mathrm{C}'$ (see Fig. \ref{fig:fourpart}).  This results in population of the state $\ket{\psi_{\mathrm{probe}}}$, where
\begin{eqnarray}
\ket{\psi_{\mathrm{probe}}} & \propto & \sum_{P=\pm 1}\bra{\mathrm{C}',J\! =\! 1,M\! =\! 0,P}\hbox{\boldmath{$\epsilon$}}_{probe}\cdot\mathbf{r}\ket{\psi_{P-}(T)} \nonumber \\
&\propto & e^{-i\delta_Z T}\bra{\mathrm{C}',J\! =\! 1,M\! =\! 0,P\! =\! +1} x \ket{J\! =\! 1,M\! =\! +1,P\! =\! -1} \nonumber \\
&+& e^{i\delta_Z T}\bra{\mathrm{C}',J\! =\! 1,M\! =\! 0,P\! =\! +1} x \ket{J\! =\! 1,M\! =\! -1,P\! =\! -1} \nonumber \\
&+& e^{-i\delta_Z T}\bra{\mathrm{C}',J\! =\! 1,M\! =\! 0,P\! =\! -1} x \ket{J\! =\! 1,M\! =\! +1,P\! =\! +1} \nonumber \\
&-& e^{i\delta_Z T}\bra{\mathrm{C}',J\! =\! 1,M\! =\! 0,P\! =\! -1} x \ket{J\! =\! 1,M\! =\! -1,P\! =\! +1} \nonumber \\
& \propto & \cos{(\delta_Z T)}\ket{\mathrm{C}',J\! =\! 1,M\! =\! 0,P\! =\! +1} \nonumber\\
&& + i\sin{(\delta_Z T)} \ket{\mathrm{C}',J\! =\! 1,M\! =\! 0,P\! =\! -1}.
\end{eqnarray}

%\begin{figure}
%\includegraphics{Cefield}
%\caption{(color online) Optical double resonance probe scheme for %oriented molecular states.  The oriented molecular states are of mixed %parity, and the linewidth of the laser driving this transition is much %broader than the $\Omega$-doublet splitting in the $\mathrm{C}'$ %state.  Hence, both $P\! = \!+1$ and $P\! = \! -1$ sublevels of the %$\mathrm{C}'$ state are coherently populated by the probe laser.
%\label{fig:Cefield}}
%\end{figure}

At first glance, it may appear that the information about the spin alignment precession cannot be retrieved from this state, since the population in the $\mathrm{C}'$ state $P_{\mathrm{C}'} \propto |\langle\psi_{\mathrm{probe}} \ket{\psi_{\mathrm{probe}}}|^2$ is independent of $T$.  However, because the angular distribution of fluorescence from the $P=+ 1$ and $P=-1$ states of $\mathrm{C}'$ is different, quantum beats are indeed present in the fluorescence.  A calculation of the beat contrast $c$ similar to that in section III shows that in this case we expect $c \approx 0.16-0.33$, with the exact value again dependent on the unknown branching ratios for $\ket{\mathrm{C}',J=1,P=-1}\rightarrow \ket{\mathrm{X},J=0,2,P=+1}$ decays (see Fig. \ref{fig:fourpart}). A brief attempt to observe this type of signal was unsuccessful, but consistent with this expectation given the experimental conditions at the time.

The quantum beat contrast for this type of signal (and all others discussed earlier) can be substantially improved by selecting a single polarization in the detected fluorescence.  Suppose that only fluorescence with polarization $\hbox{\boldmath{$\epsilon$}}_{fl} = \hat{x} (\hat{z})$ is detected.  Then all $\Delta M\! =\! 0 \left( \Delta M\! =\! \pm 1 \right)$ decays are eliminated from the sum in Eq. \ref{eq:quantumbeats} and its analogues for the other cases discussed.  In the present case of quantum beats following probe excitation, detection of only $\hbox{\boldmath{$\epsilon$}}_{fl} = \hat{z}$ results in a signal with contrast $c= 100\%$.  This can be seen easily: the signal arising from population in the $\ket{\mathrm{C}',J\! =\! 1,M\! =\! 0,P\! =\! +1}$ state is out of phase with signal from the $\ket{\mathrm{C}',J\! =\! 1,M\! =\! 0,P\! =\! -1}$ state, but the state $\ket{\mathrm{C}',J\! =\! 1,M\! =\! 0,P\! =\! +1}$ can decay \textit{only} to states $\ket{\mathrm{X},J\! =\! 1,M\! =\! \pm 1,P\! =\! -1}$.  (See Fig. \ref{fig:fourpart}.)  Surprisingly, we have found (through both modeling and measurement) that the fluoresence polarization is preserved with high fidelity even after transmission through the cylindrical quartz light pipes.  Hence, it should be possible to obtain quantum beat signals with unit contrast using this method.  In the current cell, the fluorescence polarization is scrambled by the birefringence of our sapphire windows, which leads to widely varying retardances over the angles of collected fluorescence.  We are now inserting non-birefringent (YAG) windows and plan to look for polarization-analyzed signals in the near future.

%\begin{figure}
%\includegraphics[width=3.375in]{nobeatCstructpaper}%
%\caption{(color online) Decays from the $\mathrm{C}'$$\ket{J\! =\! %1,M\! =\! 0,P\! =\! \pm 1}$ superposition state created by probing %oriented molecular states.  Decays shown in dots (red) have one phase, %and dashes (blue) are $\pi$ out of phase.  By collecting only %\hat{z}$-polarized fluorescence, only the decay channels shown in bold %($\ket{\mathrm{C}', M=0}\rightarrow \ket{X, M=0}$) will contribute to %the signal.  Since these two decay channels have the same phase, %quantum beats with high contrast ($c \cong 1$) should be %observed.\label{fig:nobeatCstructpaper}}
%\end{figure}

We envision using this detection method with a CW probe beam, since pulse-to-pulse fluctuations in the intensity of a pulsed probe beam would likely preclude shot noise limited measurements.  Emerging technologies (diode laser oscillators and fiber and/or semiconductor amplifiers) make it plausible to deliver sufficient power ($P \gtrsim 10$ W) to saturate the a-$\mathrm{C}'$ transition within $\tau_a$.  Under this assumption, we project a $\sim\!100$-fold improvement in sensitivity to $d_e$.  Potential new sources of noise (e.g. due to laser intensity fluctuations) and new systematic effects (e.g. due to AC Stark shifts induced by the probe laser) appear to be controllable at the desired levels.  We plan to implement this scheme in the near future.

\section{Microwave Absorption Probe of Spin Alignment}
%\subsection{idea of microwave absorption}

Even under ideal conditions, the optical probe scheme described above suffers from the limited solid angle of fluorescence collection available ($\lesssim 10\%$ in our apparatus).  In our previous work (Ref. \cite{firstPbO}), we pointed out that detection by monitoring absorption of a probe beam could in principle lead to much higher efficiency and sensitivity than any fluorescence-based scheme.  The condition for this improvement is that the cell have column density of $\sim\!1$ for the probe transition.  Unfortunately, there are no known electronic states of PbO which couple to the a(1) state strongly enough to meet this condition \cite{firstPbO}.  However, our success in microwave state preparation led us to consider using the $J\! =\! 1 \rightarrow J\! =\! 2$ rotational transition (a fully allowed electric dipole transition) for absorptive probing of the $J\! = \! 1$ state spin alignment.  This can be achieved by monitoring the time-dependent absorption of $\hat{x}$-polarized microwaves tuned to the $\ket{\psi_{N\pm}} \rightarrow \ket{J\!=\!2,M\!=\!0, P\! =\! +1}$ transition, in a manner exactly analogous to the optical probe described above.  Note however that the final state of definite parity is easily resolved with the microwave probe because of its dramatically smaller Doppler broadening; hence there is no complication of the type discussed for optical probing of oriented states.

In order to demonstrate the concept of this microwave probe method, we prepared an experiment to measure the a(1) state population by monitoring microwave ``absorption'' in our apparatus.  (In practice, we actually measured stimulated emission rather than absorption.)  Unfavorable Clebsh-Gordan coefficients suppress the transition strength of the spin-alignment sensitive $\Delta M = \pm 1$ transitions; hence, for this proof of concept we instead worked on the stronger $\Delta M = 0$ transitions by using $\hat{z}$-polarized microwaves.  The states $\ket{J\!=\!2,M,P\! =\! +1}$ were populated by laser excitation of the $R1$ line $(\ket{\mathrm{X},J=1,P=-1} \rightarrow \ket{\mathrm{a},J=2,P=+1})$ with $\hat{z}$ polarization, then probed on the  $\ket{J\!=\!2,M,P\! =\! +1}\!\rightarrow\!\ket{J'\!=\!1,M, P'\! =\! -1}$ transition.  The apparatus was similar to that used in the microwave state preparation experiments, but here the transmitted microwave power was monitored as a function of time following the excitation laser pulse, using a crystal detector (HP 422a).  Electrical transients from the pulsed laser were subtracted off by periodically monitoring the signal with the microwaves far off resonance.
%We define the absorption $A$ as $A = 1 - (P_a/P_0)$, where $P_a$ is the transmitted power integrated for $T = xxx \mu$s following the laser pulse, and $P_0$ is the transmitted power integrated for the same time just before the laser pulse.  \textbf{Paul: please supply alternate text for the above, describing FM detection also} Typical data for $A$ as a function of the microwave probe frequency are shown in Fig. \ref{fig:absorption}.  \textbf{Paul: this figure needs an absolute scale on the y-axis.  Please supply caption indicating detuning relative to what, and integration (single shot) and averaging time (how many shots?) for this data.} The central frequency $\omega_{12}$ of the transition was found to be $\omega_{12} = 2 \pi \times 28.217?(??)$ GHz, in good agreement with earlier, less precise measurements \cite{Martin}.\textbf{Paul: please fill in ? above} 
%Note that this measurement was performed with %$\mathcal{E} = 0$, where we expect the transition %linewidth to be dominated by Doppler broadening.  %In fact, the measured FWHM $\Gamma = 2 \pi \times %44(2)$ kHz of the absorption peak is in good %agreement with the expected value $\Gamma_D =  2 %\pi \times 42.5$ kHz at our cell temperature of %700$^\circ$ C.

\begin{figure}
\includegraphics[width=3.3in]{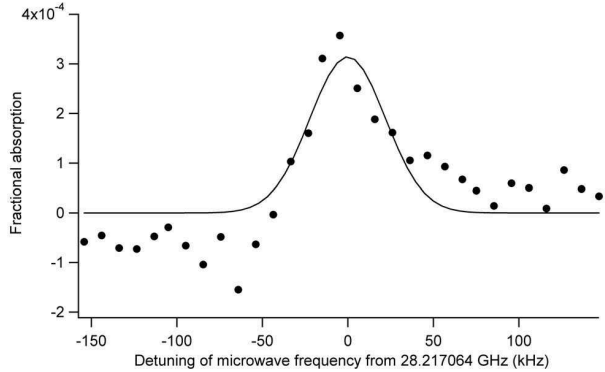}
\caption{Detection of $\ket{\mathrm{a},J=1}$ state molecules by microwave absorption on the $\ket{\mathrm{a},J=1}\rightarrow \ket{\mathrm{a},J=2}$ transition (with $\mathcal{E} = 0$). The plot shows peak fractional absorption of the microwave power as a funtion of microwave frequency.  Points are data (averaged over 512 laser pulses); the line is a fit to a Gaussian.  For this data, the fitted amplitude is $A_\mu = 3 \times 10^{-4}$; the width is $\Gamma = 2 \pi \times 52$ kHz FWHM, in reasonable agreement with the expected Doppler width $\Gamma_D =  2 \pi \times 42.5$ kHz.
\label{fig:absorption}}
\end{figure}

%\subsection{results of current microwave absorption}

As seen in Fig. \ref{fig:absorption}, the peak microwave absorption is only $A_\mu \approx 3\times 10^{-4}$ in our current configuration, even on this relatively strong probe transition.  This is not sufficient for a competitive measurement of $d_e$ using the microwave beam to probe the spin alignment precession.  However, we believe that a few modifications to the apparatus could change this situation dramatically.  For an absorptive probe measurement of this type, a long cylindrical vapor cell (rather than the roughly cubic cell now used) could increase the absorption column density.  In addition, it should be possible to add a microwave Fabry-Perot cavity around this longer cell, with partially reflecting mirrors designed to provide a finesse $\mathcal{F}\gg 1$, further increasing the effective column density.  A detailed analysis of the signal-to-noise available with this technique is beyond the scope of this paper.  However, we find that the primary noise source is likely to be thermal noise on the microwave detection amplifier; and that a factor of 1000 or more improvement in sensitivity to $d_e$ could be obtained with this detection method, relative to the fluorescence quantum beat method used at present.

We note in passing that this microwave absorption probe method also enabled a variety of useful auxiliary measurements.  For example, the transition amplitude and hence the absorption cross-section can be calculated accurately from the known molecular electric dipole moment $\mu_a$.  Hence, the absorption signal gives information on the absolute density of a(1)($J\! =\! 1$) state molecules in the vapor cell, following laser excitation.  Similar measurements on X state rotational transitions likewise yielded both the number of ground state molecules in the cell and the number excited into the a(1) state by the laser.  These results are discussed in Appendix A.

%\subsection{promise of better microwave absorption}

\section{Conclusions}

In conclusion, we have developed several techniques for the preparation of molecular states with simultaneous molecular orientation and spin alignment, as well as several methods for detecting the precession of this spin alignment.  These techniques were demonstrated with the a(1) state of PbO, as relevant for the measurement of $d_e$.  Our experiments allowed a variety of auxiliary measurements, such as the dependence of the molecular $g$-factor on the applied electric field $\mathcal{E}$.  The state preparation methods described here should enable a measurement of $d_e$ with sensitivity within an order of magnitude of the current limit, and the improved methods for detecting spin alignment precession promise measurements with substantially improved sensitivity.

We thank M. Kozlov for many useful discussions, including his derivation of the function $g(\mathcal{E})$; and F. Bay, D. Kawall, and V. Prasad for their contributions to earlier parts of the experiment. We are grateful for the support of NSF Grant No. PHY0555462.

\section{Appendix A: Absolute signal sizes and excitation efficiency}
Here we briefly outline the methods for, and results of, determining absolute signal size and excitation efficiency in our experiments.  The PbO X state vapor density is determined by microwave absorption measurements on the $\ket{X,J\! =\! 1}\rightarrow\ket{X,J\! =\! 2}$ rotational transition.  We use the Stark-modulation method \cite{TownesSchawlow}, with the modulation depth and transition strength determined by the known molecular electric dipole moment in the ground state, $\mu_X \cong 4.6$ D \cite{Herzbergconstants}.  At our nominal operating temperature of 700$^\circ$ C, we find the PbO vapor density $n_{\mathrm{PbO}}$ (after accounting for the Boltzmann distribution over rotational and vibrational levels) to be $n_{\mathrm{PbO}} \approx 3\times 10^{12}$ cm$^{-3}$.  We note that this is roughly one order of magnitude lower than the density assumed in our earlier work (Refs. \cite{2004,firstPbO}), based on the data in \cite{CRC}.  A considerable literature exists on the partial pressures of various species such as PbO, Pb$_2$O$_2$, Pb$_4$O$_4$, etc. over solid PbO in this temperature range \cite{Drowart, Popovic, Lopatin, LHildenbrad} 
%\textbf{ALL: I have listed a few of the citations here.  What key ones are missing?}.  
Although there is some disagreement among various authors on the partial pressure of PbO, our result lies within the spread of data from these earlier measurements.

We determine the efficiency of laser excitation from the X state to the a(1) state in three independent ways.  In method 1, we use microwave absorption (as above) to measure the absolute change in population of the $\ket{X,J=1}$ state following laser excitation on the $\ket{X,J=1}\rightarrow\ket{a,J=2}$ transition (R1 line).  In method 2, we measure the absolute population appearing in the $\ket{a,J=1}$ state by the microwave absorption method discussed in Section XII.  In method 3, we use the observed absolute size of our fluorescence signals, in combination with the calculated detection efficiency (see Section X), to determine the $\ket{a,J=1}$ state population.  All three methods are in agreement within a factor of $2$, and indicate that the efficiency of laser excitation is $\eta_{exc} \approx 2-4\%$.  Note also that the agreement of method 3 with the others provides independent validation of our calculated detection efficiency $\eta_{tot} = 5\times 10^{-4}$.

The expected excitation efficiency can also be calculated as follows.  As described in Ref. \cite{firstPbO}, the excitation cross-section $\sigma_{exc}$ can be determined from the known a(1) state lifetime, the estimated Franck-Condon factors, and the laser and Doppler linewidths.  (Note that the expressions for $\sigma_{exc}$ in Ref. \cite{firstPbO} contain a numerical error, and should be multiplied by a factor of 4 to obtain the correct value.)  Based on this calculation, and assuming that $\approx 1/3$ of our laser power is in a $\approx 2$ GHz frequency band, we expect $\eta_{exc} \approx 20\%$.  However, this calculation predicts that $\eta_{exc}$ should grow linearly with laser power up to our maximum available level.  Instead, we observe instead that the $\ket{a,J=1}$ state population grows roughly as the square root of the laser power, over an order of magnitude in power up to the maximum level.  We believe this is because, while the spectrum of our laser is $\approx 2$ GHz wide when averaged over many pulses, the single-shot spectrum actually consists of 1-3 much narrower lines (each as narrow as $\sim\! 30$ MHz, the Fourier transform limit of our $\sim\! 6$ ns duration pulses).  This behavior is known to be typical of pulsed dye lasers \cite{Weber1990}.  In this case, for each laser pulse the small velocity classes of the Boltzmann distribution that in resonance with each narrow spectral line can be strongly saturated, while the other classes remain un-excited.  Then, power broadening of the saturated signal leads to an increase in population proportional to the square root of the laser power.  In this model, the expected excitation efficiency is reduced by a factor of 10-20 compared to the calculation assuming a single-shot linewidth of 2 GHz.  This brings the calculated value of $\eta_{exc}$ into better agreement with our measurements.

%With the current apparatus, the detection efficiency is $\eta_d\sim7\times10^{-5}$.  The current counting rate %$\dot{N}=4\times10^7/$s and a background rate due to blackbody radiation of $\dot{B}=2\times10^{10}/$s.  In this %blackbody dominated regime the sensitivity to an EDM is
%\begin{equation}
%|d_e|=\frac{1}{W_d}\frac{4}{2 \pi c \tau_c \sqrt{N}} \sqrt{\frac{B}{S_0}}
%\end{equation}
%where $W_d=-(6.1^{+1.8}_{-0.6})\times 10^{24}$ Hz/e-cm~\cite{Titov}, $\tau_c$ is the lifetime of the state in the %cell, and $S_0$ is the signal size at $t=0$.  In this background limited case,   This leads to a statistical
%sensitivity of $\mathbf{d}_e$= $10^{-25}$ e-cm$/\sqrt{\rm{day}}$ with a 30\% measurement duty cycle.
%[numbers for cell density and laser excitation efficiency]

\section{Appendix B: Calculation of molecular matrix elements}

For many of the calculations described in main text, it is necessary to calculate lab-frame matrix elements of operators defined in the body-fixed frame of the molecule.  For example, the amplitude $A$ for the $\ket{a,J\! =\! 1,M,P\! =\! -1}\rightarrow\ket{a,J'\! =\! 2,M'\! =\! 2,P'\! =\! +1}$ microwave transition is determined by the off-diagonal matrix element of the molecular electric dipole moment operator $\hbox{\boldmath{$\mu$}}_a = \mu_a \hat{n}$
\begin{equation}
A = \bra{\mathrm{a},J'\! =\! 2,M',P'\! =\! +1}\hbox{\boldmath{$\epsilon$}}_{\mu}\cdot\hbox{\boldmath{$\mu$}}_{\mathrm{a}}\ket{\mathrm{a},J \! =\! 1,M,P\! =\! -1},
\end{equation}
where $\hbox{\boldmath{$\epsilon$}}_{\mu}$ is the microwave polarization.  Although there are a variety of texts which describe calculation of matrix elements of this type (see e.g. Refs. ~\cite{Brown2003, LandauLifshitz}), we believe the basic procedure may be unfamiliar to most atomic physicists, so we include it for completeness here.  It is most convenient to work in the basis with signed values of $\Omega$.  Consider the matrix element $M$ of a generalized irreducible tensor operator $T_{kq}$ (where $k$ is the rank and $q$ the projection) between states in this basis:
\begin{equation}
M = \bra{J',M',\Omega'}T_{kq}\ket{J,M,\Omega} = (-1)^{(J'-M')}(J'\Omega'||T_k||J\Omega)
\left( \begin{array}{ccc}
J' & k & J \\
-M' & q & M
\end{array} \right)
%\left( \stackrel{J'}{-M'}\stackrel{k}{q}\stackrel{J}{M} \right).
\end{equation}
Here we have invoked the usual Wigner-Eckhart theorem: $(J'\Omega'||T_k||J\Omega)$ is a reduced matrix element and the factor in parentheses is a $3j$ symbol.  This reduced matrix element can be expressed in terms of specific matrix elements \textit{evaluated in the molecule-fixed frame} by the transformation
\begin{equation}
(J'\Omega'||T_k||J\Omega) = \sum_{q'=-k}^{k}(-1)^{(J'-\Omega')}\sqrt{(2J'+1)(2J+1)}
\left( \begin{array}{ccc}
J' & k & J \\
-\Omega' & q' & \Omega
\end{array} \right)
%\left( \stackrel{J'}{-\Omega'}\stackrel{k}{q'}\stackrel{J}{\Omega} \right)
\bra{\Omega'}T_{kq'}\ket{\Omega}.
\end{equation}
Here $q'$ is a dummy index; at most, a single term with $q' = \Omega - \Omega'$ contributes to the sum.  The body-fixed states of the form $\ket{\Omega}$ are defined in a frame $(X,Y,Z)$ where the molecular axis $\hat{n} \equiv \hat{Z}$, and $\Omega \equiv J_{eZ}$.  The matrix elements of the body-fixed operators used in this paper ($n_z$ and $\Omega$) are trivial to evaluate in this basis.

% If in two-column mode, this environment will change to single-column

% format so that long equations can be displayed. Use


%\begin{widetext}


%\end{widetext}


\bibliography{bibliography}

\end{document}